\documentclass{emulateapj}
\usepackage{rotating}
\usepackage{rotating}

\shorttitle{M31 Dwarf Spheroidal Galaxies}
\shortauthors{Kalirai et al.}

\begin{document}


\title{The SPLASH Survey: Internal Kinematics, Chemical Abundances, and \\ 
Masses of the Andromeda I, II, III, VII, X, and XIV dSphs$^{1,2}$}


\author{
Jason S.\ Kalirai\altaffilmark{3},
Rachael~L.\ Beaton\altaffilmark{4},
Marla C.\ Geha\altaffilmark{5}, 
Karoline M.\ Gilbert\altaffilmark{6,7}, \\
Puragra Guhathakurta\altaffilmark{7},
Evan N.\ Kirby\altaffilmark{7,8,9}, 
Steven~R.\ Majewski\altaffilmark{4}, \\  
James C.\ Ostheimer\altaffilmark{4},
Richard~J.\ Patterson\altaffilmark{4}, \\ and
Joe Wolf\altaffilmark{10}
}


\altaffiltext{1}{Data presented herein were obtained at the W.\ M.\ Keck
Observatory, which is operated as a scientific partnership among the
California Institute of Technology, the University of California, and the
National Aeronautics and Space Administration.  The Observatory was made
possible by the generous financial support of the W.\ M.\ Keck Foundation.}
\altaffiltext{2}{Based on observations obtained with the Kitt Peak National 
Observatory.  Kitt Peak National Observatory of the National 
Optical Astronomy Observatories is operated by the Association of Universities 
for Research in Astronomy, Inc., under cooperative agreement with the National 
Science Foundation.}
\altaffiltext{3}{Space Telescope Science Institute, 3700 San Martin Drive, 
Baltimore MD 21218; jkalirai@stsci.edu}
\altaffiltext{4}{Department of Astronomy, University of Virginia, P.\ O.\ Box
3818, Charlottesville VA, 22903; rlb9n,srm4n,rjp0i@virginia.edu} 
\altaffiltext{5}{Astronomy Department, Yale University, New Haven CT, 06510; 
marla.geha@yale.edu}
\altaffiltext{6}{Department of Astronomy, Box 351580, University of Washington, 
Seattle WA, 98195; kgilbert@astro.washington.edu}
\altaffiltext{7}{University of California Observatories/Lick Observatory, 
University of California at Santa Cruz, Santa Cruz CA, 95064; raja@ucolick.org}
\altaffiltext{8}{Department of Astronomy, California Institute of Technology, 
Pasadena, CA, 91125; enk@astro.caltech.edu}
\altaffiltext{9}{Hubble Fellow}
\altaffiltext{10}{Center for Cosmology, Department of Physics and Astronomy, 
University of California at Irvine, Irvine CA, 92697; wolfj@uci.edu}


\begin{abstract}

We present new Keck/DEIMOS spectroscopic observations of hundreds of individual stars 
along the sightline to Andromeda's first three discovered dwarf spheroidal galaxies 
(dSphs) -- And~I, II, and III, and leverage recent observations by our 
team of three additional dSphs, And~VII, X, and XIV, as a part of the SPLASH Survey 
(Spectroscopic and Photometric Landscape of Andromeda's Stellar Halo).  Member stars 
of each dSph are isolated from foreground Milky Way dwarf and M31 field contamination 
using a variety of photometric and spectroscopic diagnostics.  Our final spectroscopic 
sample of member stars in each dSph, for which we measure accurate radial velocities 
with a median uncertainty (random plus systematic errors) of 4 -- 5~km~s$^{-1}$, includes 
80 red giants in And~I, 95 in And~II, 43 in And~III, 18 in And~VII, 22 in And~X, and 
38 in And~XIV.  The sample of confirmed members in the six dSphs are used to derive each 
system's mean radial velocity, intrinsic central velocity dispersion, mean abundance, 
abundance spread, and dynamical mass.

This combined data set presents us with a unique opportunity to perform 
the first systematic comparison of the global properties (e.g., metallicities, sizes, 
and dark matter masses) of one-third of Andromeda's total known dSph population with 
Milky Way counterparts of the same luminosity.  Our overall comparisons indicate that 
the family of dSphs in these two hosts have both similarities and differences.  For example, 
we find that the luminosity -- metallicity relation is very similar between $L$ $\sim$ 10$^{5}$ 
-- 10$^{7}$ $L_\odot$, suggesting that the chemical evolution histories of each group of 
dSphs is similar.  The lowest luminosity M31 dSphs appear to deviate from the relation, 
possibly suggesting tidal stripping.  Previous observations have noted that the sizes of 
M31's brightest dSphs are systematically larger than Milky Way satellites of similar 
luminosity.  At lower luminosities between $L$ = 10$^{4}$ -- 10$^{6}$~$L_\odot$, we find that 
the sizes of dSphs in the two hosts significantly overlap and that four of the 
faintest M31 dSphs are {\it smaller} than Milky Way counterparts.  The first dynamical 
mass measurements of six M31 dSphs over a large range in luminosity indicates similar 
mass-to-light ratios compared to Milky Way dSphs among the brighter satellites, and smaller 
mass-to-light ratios among the fainter satellites.  Combined with their similar or larger sizes 
at these luminosities, these results hint that the M31 dSphs are systematically 
less dense than Milky Way dSphs.  The implications of these similarities and differences 
for general understanding of galaxy formation and evolution is summarized.


\end{abstract}





\section{Introduction} \label{intro}

Dwarf spheroidal (dSph) galaxies represent important laboratories for constraining 
feedback processes in galaxies and for testing cosmological models 
on the smallest scales (Klypin et~al.\ 1999; Moore et~al.\ 1999; Bullock, Kravtsov, 
\& Weinberg 2000).  These low luminosity systems 
($-1.5 \lesssim M_V \lesssim -13$) show no evidence of ongoing star formation, 
contain little or no interstellar matter, and are almost always distributed 
around massive hosts, such as the Milky Way.  Within the framework of the 
$\Lambda$CDM paradigm, theories suggest that the observed dSph satellites 
are the present-day counterparts to previously accreted objects that were tidally 
destroyed to build the spheroids and stellar halos of massive galaxies 
(Bullock et~al.\ 2001; Bullock \& Johnston 2005).  This process has been 
directly verified in the outskirts of giant galaxies with the discovery 
of luminous substructure in the form of tidal streams.  For example, 
the accretion mechanism, where these dwarf galaxies are shredded and 
eventually melded into diffuse stellar components, is abundant in the 
Milky Way (e.g., the Sagittarius, Orphan, and Cetus Polar Streams -- Ibata, 
Gilmore, \& Irwin 1994; Belokurov et~al.\ 2007; Newberg, Yanny, \& Willet 2009), 
in the Andromeda (M31) spiral galaxy (e.g., the Giant Southern Stream -- 
Ibata et~al.\ 2001), as well as in several other systems such as NGC~5907 
(Martinez-Delgado et~al.\ 2008) and NGC~4013 (Martinez-Delgado et~al.\ 2009).

Despite representing the most common morphological class among all galaxies 
in the Universe, the overall count of known dSph satellites around the Milky 
Way and M31 are significantly lower than first-order expectations.  Simulations 
predict large galaxy halos to host {\em thousands} of dark matter substructures 
with masses similar to those of dSph galaxies (Klypin et~al.\ 1999; Diemand, Kuhlen, 
\& Madau 2007; Springel et~al.\ 2008).  While models exist for explaining the mismatch, 
they can only be tested with precise mass measurements of the observed dwarfs.  
Recently, large observational efforts have successfully uncovered 
many new dSphs around both the Milky Way (e.g., Willman et~al.\ 2005; 
Zucker et~al.\ 2006a; 2006b; Sakamoto \& Hasegawa 2006; Belokurov et~al.\ 2006; 
Belokurov et~al.\ 2007; Irwin et~al.\ 2007; Walsh, Jerjen, \& Willman 2007; 
Liu et~al.\ 2008; Belokurov et~al.\ 2008; Belokurov et~al.\ 2009) and M31 (Martin et~al.\ 2006; 
Zucker et~al.\ 2007; Ibata et~al.\ 2007; Majewski et~al.\ 2007; Irwin et~al.\ 2008; 
McConnachie et~al.\ 2008; Martin et~al.\ 2009), effectively doubling the total sample 
size in the past few years.  Follow up spectroscopic observations of individual 
stars in the new Milky Way satellites (e.g., by Kleyna et~al.\ 2005; Munoz et~al.\ 
2006; Simon \& Geha 2007, Martin et~al.\ 2007; Koch et~al.\ 2009; Geha et~al.\ 2009) have established 
that these systems are the most dark matter dominated objects known, 
and also provided a means to test proposed solutions of the ``missing satellites 
problem'' (see also Strigari et~al.\ 2008a; 2008b).  Unfortunately, it is difficult to draw 
general conclusions based on the Milky Way system alone.


\begin{table*}
\begin{center}
\caption{}
\begin{tabular}{llccccc}
\hline
\hline
\multicolumn{1}{c}{Date} & \multicolumn{1}{c}{Mask} & 
\multicolumn{2}{c}{Pointing center:} & \multicolumn{1}{c}{Field PA} & 
\multicolumn{1}{c}{Exp.} & \multicolumn{1}{c}{No. Sci.}  \\ 
& \multicolumn{1}{c}{} &\multicolumn{1}{c}{$\alpha_{\rm J2000}$} &
\multicolumn{1}{c}{$\delta_{\rm J2000}$} &\multicolumn{1}{c}{($^\circ$E of N)} & 
\multicolumn{1}{c}{Time} & \multicolumn{1}{c}{Targets\tablenotemark{1}} \\ 
& \multicolumn{1}{c}{} & \multicolumn{1}{c}{($\rm^h$:$\rm^m$:$\rm^s$)} &
\multicolumn{1}{c}{($^\circ$:$'$:$''$)} & & \multicolumn{1}{c}{(s)} & \\
\hline
2005 Nov 05  & d1\_1 (And~I)    & 00:45:48.61  &  +38:05:46.8 & $+0.0$  & 3 x 1,200 & 153   \\
2006 Sep 16  & d1\_2 (And~I)    & 00:46:13.95  &  +38:00:27.0 & $+90.0$ & 3 x 1,200 & 150   \\
2005 Sep 06   & d2\_1 (And~II)  & 01:17:07.46  &  +33:29:25.1 & $-90.0$ & 3 x 1,200 & 139  \\
2005 Sep 06   & d2\_2 (And~II)  & 01:16:43.29  &  +33:34:25.8 & $+0.0$  & 3 x 1,200 & 139  \\
2005 Sep 08   & d3\_1 (And~III) & 00:36:03.83  &  +36:27:27.4 & $+90.0$ & 3 x 1,200 & 128  \\
2005 Sep 08   & d3\_2 (And~III) & 00:35:39.61  &  +36:21:41.8 & $+0.0$  & 3 x 1,200 & 126  \\
\hline
\end{tabular}
\tablenotetext{1}{10 of the stars in And I, 17 in And II, and 
15 in And III were observed on both masks.}
\label{table:masks1}
\end{center}
\end{table*}

Recently, a number of large surveys have targeted the outskirts of M31 
with wide field imagers on 4-meter class telescopes (e.g., Ferguson et~al.\ 2002; 
Ostheimer 2003; Ibata et~al.\ 2007; McConnachie et~al.\ 2009).  These studies, 
combined with spectroscopic 
follow up such as the SPLASH Survey (Spectroscopic and Photometric Landscape of 
Andromeda's Stellar Halo), have resulted in ground-breaking discoveries related to 
the properties of M31's stellar halo, which itself was discovered in 
the surveys \citep{guh05,irwin05}.  For example, the field halo of M31 
exhibits a power-law surface brightness profile similar to the Milky Way 
\citep{ostheimer03,guh05,irwin05}, is filled with substructure 
\citep{ibata07,gilbert07,gilbert09}, and is metal-poor in its outskirts 
\citep{kalirai06a,chapman06,koch08}.  These recent surveys have also led to the 
discovery of new dSph galaxies in M31, 
over 35 years after \cite{vandenbergh72} found the first such systems.  
Similar to the discovery rate in the Milky Way, the past five years have 
led to thirteen new galaxies, And~IX \citep{zucker04}, And~X \citep{zucker07}, 
And~XI--XIII \citep{martin06}, And~XIV \citep{majewski07}, And~XV--XVI 
\citep{ibata07}, And~XVII \citep{irwin08}, And~XVIII--XX \citep{mcconnachie08}, 
and And~XXI--XXII \citep{martin09}.\footnote{And~XVIII is a distant background Local 
Group dSph.}  The total population of known dSphs is therefore 
similar in M31 and the Milky Way, and considering the spatial coverage of the 
surveys in both galaxies (e.g., the PAandAS Survey -- Ibata et~al.\ 2007; 
McConnachie et~al.\ 2008; McConnachie et~al.\ 2009), the overall census is 
incomplete by at least a factor of a few.

Given the distance to M31, 780~kpc, most of our information on the properties 
of its dSph population has come about from studies of the bright red giant branch (RGB) 
stars in these systems.  For example, \cite{mcconnachie06} present a detailed analysis 
of the structural properties of six of these galaxies and find both similarities 
and differences with the Galactic population.  For example, similar to the 
Galactic system, M31 dSphs with higher central surface brightnesses  are found 
further from their host.  A key difference however is that the scale radii 
of the M31 dSphs is approximately twice as large as the Milky Way dSphs at the 
same luminosity. \cite{mcconnachie06} also show that the central surface brightness 
of the M31 satellites is smaller than that of comparable Milky Way systems.  The 
only Hubble Space Telescope {\it HST} observations of the M31 satellites are the 
WFPC2 studies of Da Costa et~al.\ (1996; 2000; 2002), who obtained optical 
photometry of stars in each of And~I, II, and III down to the level of the horizontal 
branch.  Their study provides the first constraints on the age and abundance 
distributions of stars in these three systems.  Indeed, Da Costa et~al.\ find 
that, similar to the Milky Way dSphs, the M31 satellites have also had diverse 
evolutionary histories.

Unlike for the Milky Way satellites, there exist very few {\it spectroscopic} 
studies of the M31 dSph system, most of which were only able to characterize a 
few stars in each galaxy.  \cite{cote99b} obtained the first such spectra 
using HIRES on Keck.  They measured velocities for seven RGB stars in And~II and 
estimated both the radial velocity of the galaxy and the central velocity dispersion 
from this small sample.  \cite{guh00} also obtained spectra of a few 
tens of individual RGB stars in each of And I, III, V, and VI.  Their study led 
to a dynamical mass estimate of M31 based on the mean radial velocity of each 
dSph \citep{evans00}. Unfortunately, the data set could not be used to probe the 
internal kinematics of the satellites given low numbers of stars in each galaxy, 
and large uncertainties in the velocity measurements.  \cite{guh00} also looked at 
several stars in And~VII with HIRES on Keck and measured both the radial velocity and 
the total velocity dispersion of the galaxy (e.g., the combined true intrinsic 
dispersion plus the dispersion due to velocity errors).  Recently, \cite{chapman05} 
and \cite{chapman07} have used Keck/DEIMOS to spectroscopically confirm six to eight 
members of And~IX and And~XII, however these data do not constrain the internal kinematics 
of the dSph.  For example, the inclusion of one marginal member in the outer radial 
bin of And~IX inflates the velocity dispersion by a factor of two.  A re-analysis of these 
data by \cite{collins09} establishes upper limits on the velocity dispersions of both 
galaxies\footnote{At the time of the writing of this paper, the \cite{collins09} study 
appears as a preprint.  The 1-$\sigma$ error bars on the velocity dispersions are consistent 
with $\sigma_v$ = 0~km~s$^{-1}$ for both dSphs.}, and they also do not resolve the dispersions of And~XI and XIII.  
Finally, \cite{letarte09} have presented a new analysis of seven 
and eight spectroscopically confirmed stars in And~XV and And~XVI.  The radial 
velocity uncertainties in these measurements are large, ranging from $\sigma_v$ = 6 -- 
25~km~s$^{-1}$, with one star as high as $\sigma_v$ = 52~km~s$^{-1}$.  Summarizing, 
of the entire 18 galaxies comprising the known M31 dSph 
population, And~II and And~VII are the only systems with secure velocity dispersion 
measurements, based on high resolution studies of 7 and $\sim$20 stars 
each.

In this paper, we present the first results from a large project aimed at 
spectroscopically observing hundreds of stars in the M31 dSph system, as a part of 
the overall SPLASH Survey.  These data will allow the first detailed comparison of 
the properties of M31 satellites to the Milky Way dSph system.  For example, we 
will accurately measure velocities of (a minimum of) dozens of individual stars 
in each galaxy, thereby resolving their intrinsic central velocity dispersions and 
leading to estimates of the total system masses.  For the best data sets ($>$500 
velocities in a single dSph), we will also probe for any radial changes in the 
velocity dispersion, and test for rotation, thereby constraining dynamical models 
for dSphs (e.g., Kroupa 1997).  Such high quality data sets can also be used to 
better understand how luminosity maps to dark matter subhalo mass.

In addition to the kinematical analysis, spectroscopic observations of large numbers of 
stars in each dSph can provide a detailed analysis of radial trends in the chemical abundances 
of stars, and perhaps even constrain the $\alpha$-abundance of these satellites.  Therefore, 
these observations directly relate to a better understanding of both population gradients 
and the evolutionary history of dSphs.  Comparisons of the results from the survey with 
the well studied dSphs of the Milky Way will aid in our understanding of whether the 
observed differences in the satellite systems (see \S\,\ref{globalproperties}) are caused 
by in situ processes within the galaxies, or whether they are shaped by different 
interaction histories of the dSphs with their massive hosts. 

In this first paper, we address the internal kinematics, chemical compositions, sizes, 
and total masses of And I, II, III, VII, X, and XIV, based on a large sample size of 
individual radial velocities.  Future papers will 
address other M31 dSphs, for which we have already collected a substantial amount of 
Keck/DEIMOS observations.  The layout of this paper 
is as follows.  We describe the photometric and spectroscopic observations of 
these dSphs as well as the data reduction in the next section.  This 
includes a detailed discussion of the uncertainties in the velocity measurements, 
which are critical to understand in order to resolve the intrinsic dispersions 
of the (kinematically cold) satellites.  Membership of stars in the satellites 
is established and velocity histograms of the confirmed red 
giants are presented in \S\,\ref{membership}.  The kinematics are analyzed 
to yield mass-to-light ratios for the galaxies in \S\,\ref{meanvelocity}.  
Next, we measure the chemical abundance of each dSph in \S\,\ref{chemabund}.  Unlike 
previous studies, we measure the abundances of each confirmed dSph star using two 
independent methods, photometrically and spectroscopically.  The results from this 
overall study are discussed in \S\,\ref{globalproperties}, where we compare the 
properties of all six M31 dSphs (e.g., their sizes, metallicities, and masses) with 
measured trends for Milky Way satellites.  The main results from the paper are summarized 
in \S\,\ref{conclusion}.


\begin{figure*}
\begin{center}
\leavevmode
\includegraphics[width=5.4cm,angle=270]{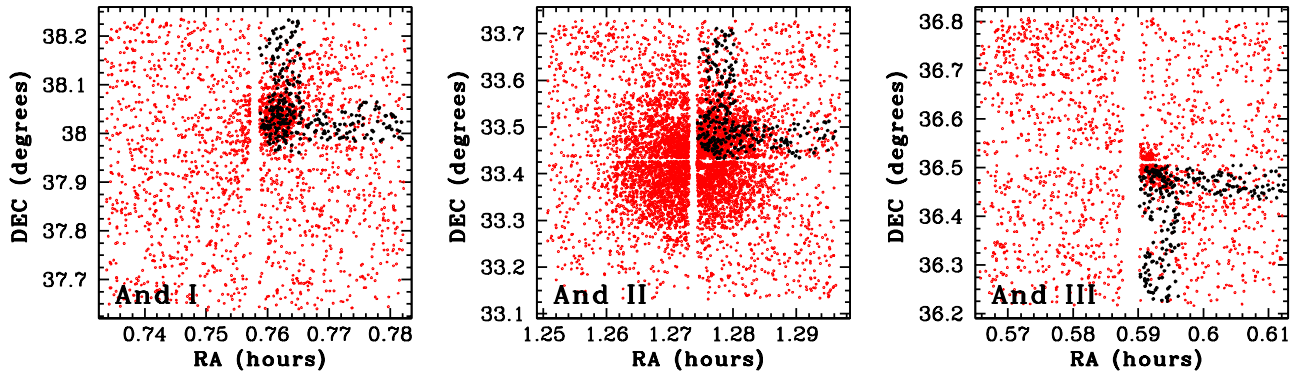}
\end{center}
\caption{The distribution of stars from the KPNO wide-field Mosaic observations 
of each dSph are shown as small points.  The selection function used to 
isolate these stars involved a simple cut on the color-magnitude diagram (CMD) 
to enhance the dSph RGB with respect to the field population, which is still 
abundantly present in the maps.  Superimposed on the photometric data are the objects 
targeted with Keck/DEIMOS spectroscopy (darker points).}
\label{fig:radecall}
\end{figure*}


\section{The Data Set}\label{datasample}

The imaging and spectroscopic observations for three of the six dSphs in this study 
are being presented for the first time (And~I, II, and III), whereas the data 
sets for And VII, X, and XIV have been (at least partially) presented in previous 
studies.  For these three galaxies, however, we present a new analysis of the 
spectra in the present paper.  We begin by first introducing the new data for 
And~I, II, and III, followed by a short summary of the previously presented data 
for And VII, X, and XIV, including any relevant new analysis.

\subsection{Photometric and Spectroscopic Observations -- And~I, II, and III}\label{observations}

We imaged each of And I, II, and III with the wide-field Mosaic camera on 
the Kitt Peak National Observatory (KPNO) 4-m telescope from 1998 to 2002 (Ostheimer 2003).  
This camera subtends an angular size of 36$' \times$ 36$'$, much larger than the 
King limiting diameter of And~I ($r_{l}$ = 10.4$'$) and And III ($r_{l}$ = 7.2$'$), 
and slightly smaller than that of And~II ($r_{l}$ = 22.0$'$).  The observations 
were obtained in the Washington System ($M$, $T_2$) bands and converted to 
Johnson-Cousins ($V$, $I$) using the relations 
given in \cite{majewski00}.  The exposure times were set at 900 seconds in the 
$M$ filter and 3,600 seconds in the $T$ filter, reaching a photometric depth of 
$I \sim$ 24.  The most luminous (metal-poor) RGB stars have intrinsic magnitudes 
of $M_{I}$ $\sim$ $-$4, and therefore at the distance of M31 (780~kpc, 
($m-M$)$_{\rm 0}$ = 24.5), these stars will have $I_{\rm 0}$ = 20.5.  Our imaging 
observations are therefore well suited to probe several magnitudes of the RGB in each of the 
dSphs.  

We also obtained imaging observations of the same fields using the intermediate-width 
$DDO51$ filter \citep{majewski00}. This filter is centered on the surface-gravity 
sensitive Mgb and MgH stellar absorption features, and therefore provides an 
efficient means to discriminate foreground Milky Way contaminants from M31 dSph 
RGB stars \citep{ostheimer03,gilbert06,guh06}.  The filter is especially efficient at 
boosting our odds of targeting true dSph stars that are located far from the core of 
each galaxy, where the surface brightness of members if low relative to the foreground 
veil of Milky Way dwarf stars.  These stars, in the outskirts of the 
dSphs, represent key data points when studying radial trends.  The combination 
of the wide-field broadband imaging observations with co-spatial $DDO51$ filter 
exposures provides the necessary framework to conduct the first largely 
uncontaminated study of the properties of M31 dSphs.

With this unique photometric data set in hand, we spectroscopically targeted 
the brightest RGB candidates in each of And I, II, and III with the DEIMOS 
multiobject spectrograph \citep{faber03} on the Keck~II 10-meter telescope.  
To do this, we first measured the positions of all stars in the KPNO Mosaic 
images and transformed the positions into equatorial coordinates using the 
USNO-B Guide Star Catalogue as a reference 
\citep{monet03}.  These positions were then mapped onto the DEIMOS mask.  
The spectroscopic data were collected in late 2005, as summarized in Table~1.  
The observations were obtained using the standard setup described in detail in several 
previous papers (e.g., Gilbert et~al.\ 2006, Guhathakurta et~al.\ 2006 and Kalirai 
et~al.\ 2006b).  To 
summarize a few key details, the spectroscopic observations were obtained 
using the 1200~lines mm$^{-1}$ grating at a central wavelength of 7800~${\rm \AA}$.  
The resolution of the spectra for a typical seeing of 0$\farcs$8 FWHM is 1.3~${\rm \AA}$ 
and the spectral range is $\approx$6400 -- 9100~${\rm \AA}$, depending exactly 
on the position of each object relative to the mask center.

In total, we measured a spectrum for 835 objects in six masks (two per galaxy), 
where each mask subtends roughly 16$' \times$ 4$'$.  Of these, 42 objects were 
measured on multiple masks to assess our radial velocity accuracy (see 
\S\,\ref{Aband}).  The targets were chosen over a magnitude range extending 
from just above the tip of the RGB down to $\sim$2 magnitudes below the tip 
in each satellite.  The mean spectral $S/N$ in the entire sample of observations 
is $\sim$5 per pixel.  The $DDO51$ filter observations were used to screen objects in the selection 
process to ensure that high probability RGB stars were targeted before lower 
probability objects (which turn out to be mostly foreground Milky Way dwarfs).

The positions and orientations of the Keck/DEIMOS spectroscopic fields relative 
to And I, II, and III were carefully placed to allow both a study of stars near 
the center of each dSph and also to probe radially outward from the core.  To 
illustrate these pointings, we present starcount maps of each galaxy taken from 
our wide-field KPNO Mosaic imaging data in Figure~\ref{fig:radecall}.  For each 
galaxy, we make a very rough cut on the photometric data to isolate stars located 
along the RGB in the color-magnitude diagrams (CMDs), which are presented later 
in \S\,\ref{membership}.  The spectroscopic targets are shown as darker points.

\subsubsection{Existing Observations of And~VII, X, and XIV}\label{additionaldSphs}

The data sets for And~VII, X, and XIV differ from the three dSphs discussed above 
in several ways, and so we briefly summarize this sample here.  First, for And~VII, 
the photometry was obtained by Grebel \& Guhathakurta (1999) using Keck/LRIS.  The 
galaxy is one of the most luminous M31 satellites known, with an integrated brightness 
of $M_{\rm V}$ = $-$13.3 $\pm$ 0.3, and so several of the brightest red giants in the 
photometric sample were targeted with Keck/HIRES for spectroscopy.  These observations 
utilized special purpose multislit masks with a half dozen slitlets each, with each 
slitlet having a length of 1 -- 2$\farcs$5.  The technique employed is described in 
detail in \cite{sneden04}, in the context of similar observations of the globular 
cluster M3.  Note, \cite{grebel99} refer to this galaxy as the Cassiopeia dSph in 
their paper.

The second galaxy in this existing sample is And~X, a newly discovered low-luminosity 
M31 dSph from the Sloan Digital Sky Survey \citep{zucker07}.  We used the imaging data 
set described by \cite{zucker07} to target two Keck/DEIMOS spectroscopic masks in this galaxy, using the 
same setup as discussed above (and below) for And~I, II, III.  The details of the 
observations, data reduction, and results are presented in \cite{kalirai09}.  

The final galaxy in the sample is And~XIV, a system that we discovered with KPNO/Mosaic 
three years ago.  This galaxy is one of the most remote systems known in M31, located at a 
projected distance of 162~kpc and a line of sight distance that places it behind M31 by 
$\sim$80~kpc.  We targeted And~XIV with two Keck/DEIMOS spectroscopic masks, again, using 
an identical setup as discussed above (and below) for And~I, II, and III.  The details 
of the observations, data reduction, and results for this dSph are presented in 
\citep{majewski07}. 

\subsection{Data Reduction}\label{data}

Data reduction of the six DEIMOS multi-slit masks being presented for the first time 
in the present study (And I, II, and III) are based on the {\tt spec2d} software 
pipeline (version~1.1.4) developed by the DEEP2 Galaxy Redshift Survey team at the 
University of California-Berkeley for that project (see Faber et~al.\ 2007 for 
more information).  Briefly, internal quartz flat-field exposures are obtained 
and used to rectify the curved raw spectra into rectangular arrays by applying 
small shifts and interpolating in the spatial direction.  A one-dimensional slit 
function correction and two-dimensional flat-field and fringing correction are 
applied to each slitlet.  Using the DEIMOS optical model as a starting point, 
a two-dimensional wavelength solution is determined from multiple Kr-Ar-Ne-Xe 
arc lamp exposures with residuals of order 0.01~${\rm \AA}$.  Each slitlet is then 
sky-subtracted exposure by exposure using a B-spline model for the sky.  The 
individual exposures of the slitlet are averaged with cosmic-ray rejection and
inverse-variance weighting.  Finally one dimensional spectra are extracted for 
all science targets using the optimal scheme of \citet{hor86} and are rebinned 
into logarithmic wavelength bins with 15~km~s$^{-1}$ per pixel.  We note that the 
modifications to this pipeline discussed by \cite{simon07} and \cite{gilbert09} were 
used in the present reductions. Specifically, the cosmic ray rejection algorithm was altered 
to allow alignment of individual two dimensional exposures in the spatial direction 
before co-addition.

\begin{figure*}
\begin{center}
\leavevmode
\includegraphics[width=13cm,angle=270]{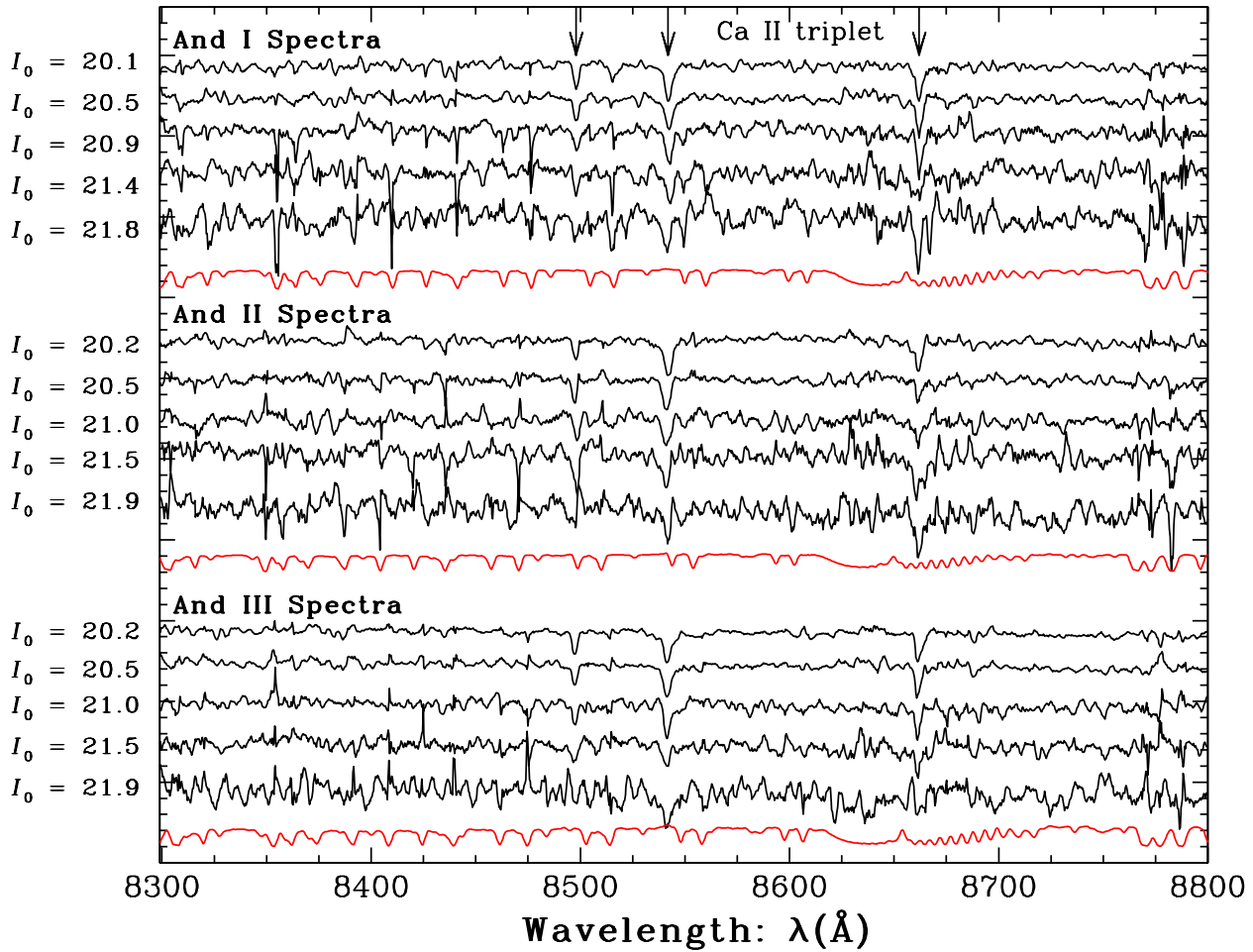}
\end{center}
\caption{Five sample spectra of RGB stars in And I, II, and III corrected to 
zero velocity.  To illustrate the quality of the full data set, these spectra 
have been chosen over a magnitude range extending from the brightest to 
the faintest stars in our spectroscopic sample.  The wavelength window of 
500~${\rm \AA}$ has been centered near the Ca\,{\sc ii} triplet absorption 
lines (8498, 8542, and 8662~${\rm \AA}$) which are clearly detected in all 
objects.  The red curve curve shows a sample inverse variance spectrum for each dSph,
calculated under the assumption that the errors are dominated by Poisson
statistics in the sky+object counts, and intended to help distinguish between
real features (e.g., Ca\,{\sc ii} triplet absorption lines) and noise artifacts 
for the faintest stars plotted.  The full spectral 
coverage of these data is $\sim$2,700~${\rm \AA}$,  which allows multiple 
features to be used in the cross-correlation to yield accurate radial 
velocities.  The spectra have been smoothed using a 7~pixel boxcar function 
for illustration purposes only.}
\label{fig:spectra}
\end{figure*}

In Figure~\ref{fig:spectra}, we present several of our final 1-dimensional spectra 
for stars in each of the three M31 dSphs.  Only a 500~${\rm \AA}$ section of the spectrum 
centered near the Ca\,{\sc ii} triplet is shown in each case for the sake of clarity.
These absorption lines, as well as other spectral features, are clearly visible in 
the data for bright and faint objects in each galaxy.


\subsection{Radial Velocity Measurements} \label{Aband}

Radial velocities are measured for all extracted one-dimensional
spectra by cross-correlating the observations with a series of
high signal-to-noise stellar templates.  The stellar templates
were observed using the same DEIMOS setup described above, and
range in spectral type from F8~III to M8~III, and also include 
subgiants and dwarfs.  For the faintest objects with lower 
signal-to-noise, this cross-correlation does not yield meaningful 
results and therefore we manually check all of the template fits and 
only include those spectra in which the template matched at least two 
different spectral features (see Guhathakurta et~al.\ 2006 for more 
details).  Of the initial sample of 835 spectra, 
this left 426 spectra after eliminating galaxies, 148 in And~I, 136 in 
And~II, and 142 in And~III.  These numbers include 1 duplicate in 
the And~I observations, 1 in And~II, and 8 in And~III.  The duplicates 
represent observations of the same star on both masks, where both spectra 
yielded a reliable velocity measurement from the cross correlation.

Below the set of sample spectra for each dSph in Figure~\ref{fig:spectra} is shown the
inverse of the variance spectrum (red), computed under the idealized
assumption that it is dominated by Poisson statistics in the sky+object
counts.  The spectral measurements presented in this paper---the
cross-correlation analysis used to measure radial velocities and the
measurement of absorption line strengths (see \S\,\ref{specmet})---properly take 
into account this dependence of the uncertainty in the spectral flux on wavelength.  
The dips in the inverse variance spectrum correspond to bright night sky emission
lines, shifted in wavelength to account for the shift in the science spectra
from the observer frame to the rest frame.  Occasionally, systematic errors
in sky subtraction can cause the actual uncertainty to be greater than the
Poisson errors shown here.  The inverse variance spectra are plotted to help
the reader distinguish between real features (e.g., the Ca\,{\sc ii} triplet 
absorption lines) and noise artifacts for the fainter stars in our sample.

The uncertainties in our velocity measurements are empirically estimated 
to be $\sim$10~km~s$^{-1}$, based on the comparison of independent 
observations from these duplicates.  While this uncertainty is many 
times less than the velocity dispersion in bulge and halo-like populations, 
it is comparable to, if not larger than, the expected velocity dispersion 
in small dSphs.  A significant fraction of the radial velocity uncertainty 
budget derives from random and systematic {\it astrometric} errors.  As 
discussed earlier, in designing our spectroscopic masks we used the 
USNO-B Guide Star Catalogue to calculate absolute positions of all stars.  The 
internal accuracy of this catalogue is known to be $\sim$0$\farcs$2 
\citep{monet03}.  At the DEIMOS pixel scale (0$\farcs$1185 per pixel), this 
uncertainty alone translates to 1.7 pixels.  Given that the dispersion of 
the 1200~lines mm$^{-1}$ grating is 0.33~${\rm \AA}$~pixel$^{-1}$ and the anamorphic 
factor is 0.6 at 8000~${\rm \AA}$, a 1.7 pixel uncertainty results in a 0.3~${\rm \AA}$ 
wavelength uncertainty.  Therefore, a random misalignment of the centroid of a star 
by 0$\farcs$2 in the slit can lead to a velocity error of 12~km~s$^{-1}$.  
The positional accuracy of a given target will also depend on the overall solution of 
the astrometric transformation and therefore depends on the density of astrometric 
reference stars 
available at a particular location in the imaging frame.  Finally, we note that 
a number of other effects can also lead to uncertainties in the astrometry.  These 
can include positional uncertainties in the images (e.g., if one or more reference 
stars have an appreciable proper motion), slight imperfections in the mechanical 
cutting process that produces the slits, and even small misalignments (i.e., 
rotations or shifts) of the entire mask during the observations.

Fortunately, these astrometric uncertainties leave a signature that can 
be used to calculate the amount that the radial velocity measurement is 
offset from the true value.  This method, as discussed by \cite{sohn07} and \cite{simon07}, 
utilizes atmospheric absorption features that are superimposed on the stellar 
spectra.  The strongest of these telluric lines are in the A-band (7600 -- 
7630~${\rm \AA}$).  To correct for the mis-centering, we observed several bright, 
rapidly rotating, hot standard stars with few emission or absorption lines 
(e.g., the Wolf Rayet star Wolf1346).  The near featureless spectra of these 
stars displays a very clean detection of the A-band absorption.  The telluric 
templates were created by allowing these stars to drift perpendicularly across 
the slit during each short exposure, ensuring that the starlight evenly fills 
the slit.  Next, we averaged the spectra for all 
such standards and cross-correlated a windowed region of the coadded spectra 
at the wavelength of the A-band absorption to the observed stellar spectra.  
For a star that is perfectly centered in the slit, this cross-correlation 
produces a near perfect match indicating that no offset is required.  However, 
for a star that is mis-centered in the slit, the observed wavelength of the 
A-band is offset relative to the standard star template.  We use this offset 
to calculate a velocity shift and apply this correction to each of our 426 
radial velocities.  The mean offset for each mask ranged from 0.7 -- 
10.2~km~s$^{-1}$, with a standard deviation of 5 -- 13~km~s$^{-1}$.  The 
final addition to each of our velocity measurements includes a heliocentric 
correction to convert our geocentric velocities into a heliocentric frame of 
reference.


\begin{figure}
\begin{center}
\leavevmode
\includegraphics[width=8.2cm]{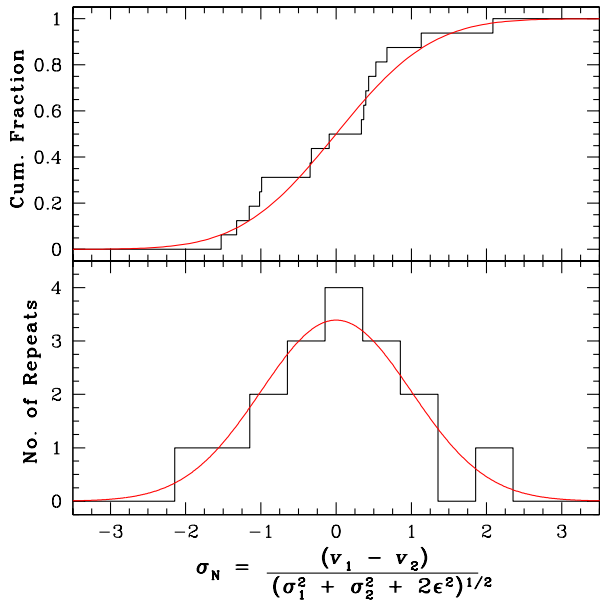}
\end{center}
\caption{Repeat measurements of the same stars in our And~I, II, III, and X 
data sets are used to define a normalized error, $\sigma_N$ equal to the 
ratio of the difference in velocity of the duplicates and the quadrature 
sum of all error terms.  In addition to the Monte Carlo error estimates, 
we find that an additional error ($\epsilon$) is needed to reproduce a 
unit Gaussian in the resulting distribution of $\sigma_N$ (solid histogram).  As 
discussed in  \S\,\ref{RVuncertainties}, we formally add $\epsilon$ = 
2.2~km~s$^{-1}$ in quadrature to each of our radial velocity errors to 
match the results from Simon \& Geha (2007), which are based on a much larger 
sample of repeat measurements taken with a similar setup to this work.}
\label{fig:dups4}
\end{figure}


\subsection{Radial Velocity Uncertainties} \label{RVuncertainties}

The observed velocity dispersions of the M31 satellites in this study reflect 
a combination of the actual intrinsic dispersions, plus the velocity errors.  
The known intrinsic dispersion of other Local Group satellites are often 
5 -- 10~km~s$^{-1}$, which is comparable to the expected errors in our 
velocity measurements.  It is therefore very important to have a good understanding 
of the true {\em uncertainties} in our measurements.  We measure these uncertainties 
using a Monte Carlo method, where we first take each stellar spectrum and add noise 
to each pixel scaled by the estimated variance in that pixel (also see discussion in 
Simon \& Geha 2007).  This procedure is repeated 1000 times assuming the variance 
in each pixel is distributed according to Poisson statistics, and the velocity 
and telluric correction is re-calculated after each run of the simulation.  The 
error in the velocity ($\sigma$) is defined as the square root of the variance in the 
recovered mean velocity.

We next compare this error measurement to the known uncertainty, measured as 
the difference in velocity between independent measurements of the same star 
(e.g., $v_1$ and $v_2$).  For this, we use the 10 duplicates with reliable 
velocity measurements in this data set, as well as the 7 duplicates in our And~X data set 
\citep{kalirai09}.  For these stars, we define a normalized error, $\sigma_N$, 
as the ratio of the velocity difference between duplicate measurements and 
the quadrature sum of all error contributions ($\sigma_1$, $\sigma_2$, and 
$\sqrt{2}$$\epsilon$).  The latter includes the Monte Carlo errors in each of the two 
velocity measurements, as well as an additional term representing any other 
error that we have not accounted for.  The distribution of $\sigma_N$ 
for all duplicates is shown in Figure~\ref{fig:dups4}, and in order to reproduce 
a unit Gaussian, we measure $\epsilon$ to be $\lesssim$3~km~s$^{-1}$.  For our 
final velocity errors, we choose to formally add an $\epsilon$ of 
2.2~km~s$^{-1}$ in quadrature with the Monte Carlo errors.  Although our measured 
value is slightly larger than this, 2.2~km~s$^{-1}$ reflects the $\epsilon$ measured 
by \cite{simon07} from a similar Keck/DEIMOS study, but with a factor of three larger 
data set of duplicate measurements.  With only 17 data points, our measured $\sigma_N$ 
is much more susceptible to a few outliers than their sample.  We note that the 
resulting velocity errors from this analysis, and with this value of $\epsilon$, have 
been directly tested by \cite{simon07} by comparing to high dispersion observations of 
a Milky Way globular cluster and dSph galaxy, each with velocity uncertainties 
of $<$1~km~s$^{-1}$.  The results from the Keck/DEIMOS analysis are in excellent 
agreement ($<$1$\sigma$) with the high resolution studies.


\begin{figure}
\begin{center}
\leavevmode
\includegraphics[width=8.5cm]{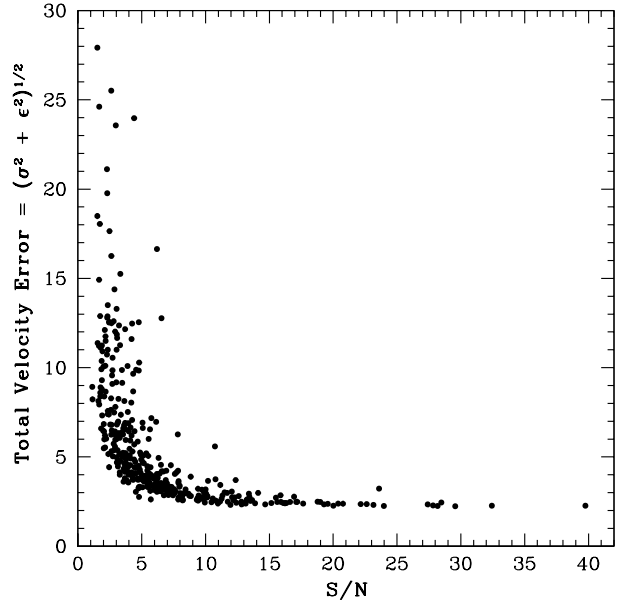}
\end{center}
\caption{The total uncertainty in our velocity measurements for all stars along the 
sightline to And I, II, and III is shown as a function of the mean spectral 
$S/N$ per pixel.  The uncertainties reflect both random and systematic velocity error contributions, 
the two being added in quadrature as discussed in \S\,\ref{RVuncertainties}.  Even 
for spectra with $S/N$ = 5, we are able to measure velocities accurately to 
$\sigma_v$ = 4~km~s$^{-1}$.}
\label{fig:s2n2}
\end{figure}


In Figure~\ref{fig:s2n2} we present the final error distribution of all velocity 
measurements in our data set, as a function of the spectral $S/N$ (per pixel).  As expected, 
the highest $S/N$ stars have a velocity error equal to the ``floor'' established 
above (2.2~km~s$^{-1}$) whereas the error increases to $\sim$4~km~s$^{-1}$ for 
stars with $S/N$ = 5, and quickly increases for lower $S/N$ spectra.  The median 
uncertainty across the full sample is 4.4~km~s$^{-1}$.  


\begin{figure*}
\begin{center}
\leavevmode
\includegraphics[width=6.5cm,angle=270]{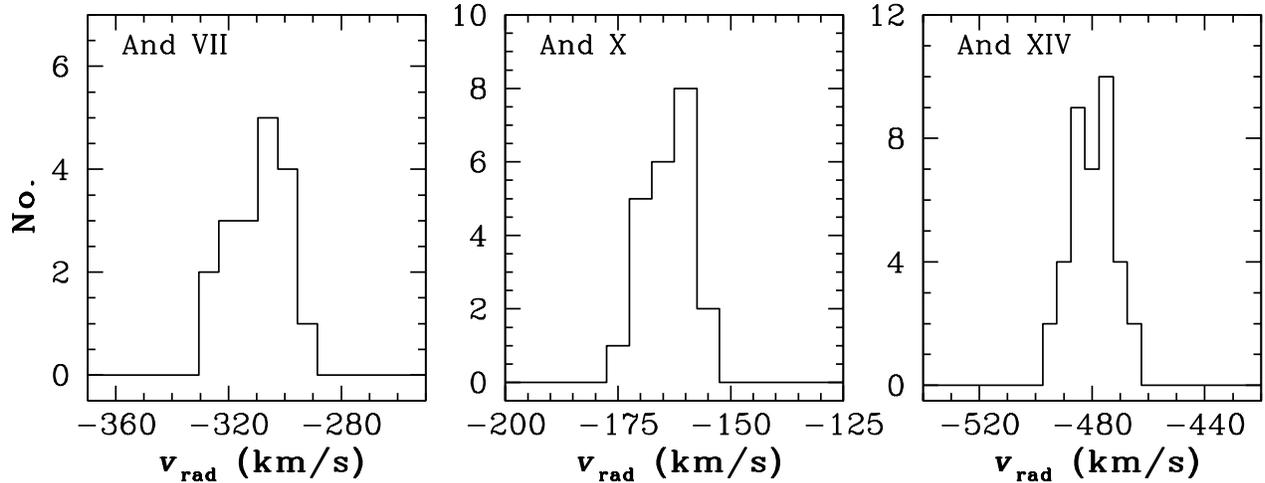}
\end{center}
\caption{The velocity histograms of And~VII ({\it left} -- Guhathakurta, 
Reitzel, \& Grebel 2000), And~X ({\it middle} -- Kalirai et~al.\ 2009), and 
And~XIV ({\it right} -- Majewski et~al.\ 2007).  The properties of each of 
these satellites, as measured from these kinematical data, are discussed in the 
text.}
\label{fig:otherdSphs}
\end{figure*}


We stress that the analysis presented above in carefully understanding the 
velocity errors in these data is essential to accurately probe the internal 
kinematics of dwarf galaxies.  The uncertainty introduced by even very small 
astrometric errors (e.g., at the 0$\farcs$05 -- 0$\farcs$1 level) in the centroids 
of targets leads to radial velocity uncertainties that are larger than the 
dispersions of some satellites.  Furthermore, these uncertainties, for a given 
spectrum, systematically offset the wavelengths of {\it all} spectral 
features by the same amount.  The rms scatter between velocity measurements, as 
measured from different absorption lines, can therefore not provide 
an accurate assessment of the overall uncertainty in radial velocity.


\begin{figure*}
\begin{center}
\leavevmode
\includegraphics[width=13cm,angle=270]{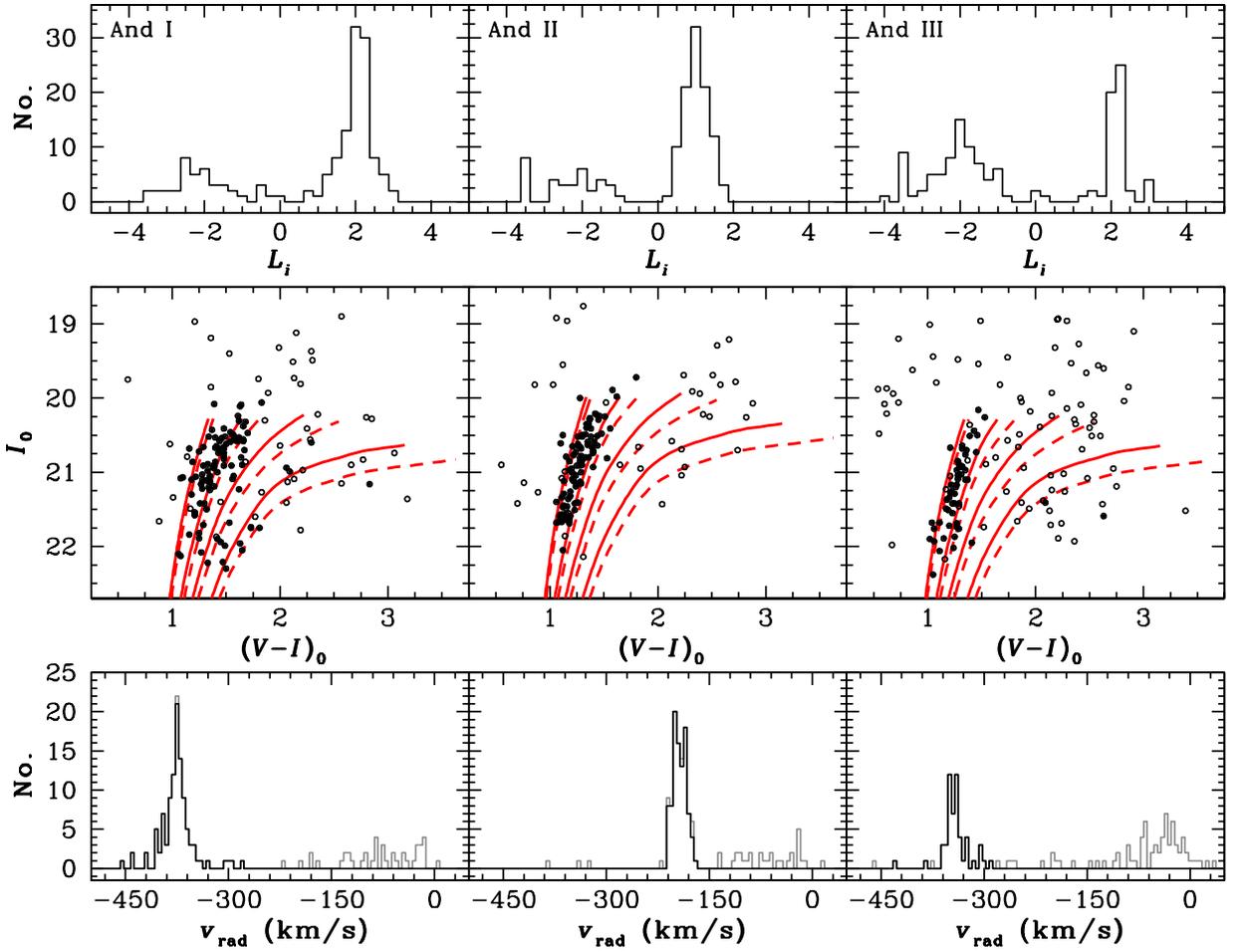}
\end{center}
\caption{{\it Top} -- The distribution of likelihoods of individual stars being 
giants at the distance of M31 ($L_i >$ 0) or foreground Milky Way dwarfs ($L_i <$ 0) 
for each of our And~I, II, and III sightlines.  The likelihoods are 
constructed using five different photometric and spectroscopic diagnostics as 
discussed in \S\,\ref{membership} and \cite{gilbert06}.  We consider the giants 
to be those stars that are three times more likely to be M31 giants as compared to 
Milky Way dwarfs ($L_i$ $\geq$ 0.5).  {\it Middle} -- CMDs for And~I, II, and III illustrating our giant 
(filled points) and dwarf (open points) stars.  The giants clearly form a tight 
RGB that follows the most metal-poor isochrones shown (see text).  The filled 
points with redder colors that do not follow the locus of most giants are 
field M31 halo stars, as discussed in \S\,\ref{membership}.  {\it Bottom} -- 
Radial velocity histograms for each of our And I, II, and III sightlines show a 
dominant cold population at large negative velocities (the dSphs) as well as a smaller 
underlying broad distribution (M31 field halo RGB stars).  Foreground Milky Way 
dwarfs are also seen at small velocities and are well separated from the dSph 
stars.  The grey histogram shows all stars for which we measured a velocity 
whereas the black histogram illustrates our giant sample (dSph members and 
M31 field halo stars).  As discussed in \S\,\ref{membership}, the And~I sightline is 
contaminated by M31 field substructure with a similar velocity to the dSph.  This 
substructure is present in this Figure, and is removed to create our 
final member dSph RGB sample as illustrated in Figure~\ref{fig:member2d1}.}
\label{fig:giant}
\end{figure*}


\subsection{Radial Velocity Measurements for \\ And~VII, X, and XIV} \label{RVuncertainties2}

The method used to reduce the And~X and XIV Keck/DEIMOS spectroscopic data, including the 
measurement of radial velocity and uncertainty, is very similar to the description above 
and has been presented in Kalirai et~al.\ 2009 and Majewski et~al.\ 2007).  For And~VII, the 
Keck/HIRES data were analyzed slightly differently.  The sky subtraction was performed 
by combining the few pixels of sky in all of the slits into a master sky spectrum.  
The velocities were extracted by cross correlating the And~VII target spectra 
with radial velocity standard stars.  The latter were observed through the same 
slitlets on the same HIRES masks as the science observations.  Although the 
$S/N$ of the spectra are quite low, the wide baseline in wavelength allowed reliable 
velocities to be measured for 18 of the 21 And~VII RGB candidates.  The uncertainties 
in the individual velocity measurements are $\lesssim$1.5~km~s$^{-1}$.

\section{Establishing Member Stars in Each dSph}\label{membership}

Membership of stars in the three previously observed dSphs, And~VII, X, and XIV, 
have been discussed in \cite{guh00}, \cite{kalirai09}, and \cite{majewski00}, and so 
we refer to those papers for the details (the analysis of And~X and XIV is identical 
to the discussion below).  Summarizing, accurate radial velocities and uncertainties 
were established for 18 member stars in And~VII, 22 member stars in And~X, and 
38 member stars in And~XIV.  These papers, as well as Grebel \& Guhathakurta (1999), 
present the CMDs for each of these satellites.  We summarize the three radial velocity 
histograms for these galaxies in Figure~\ref{fig:otherdSphs}, and now focus on the 
new observations of And~I, II, and III.

In section~\ref{intro} we noted that only one previous spectroscopic study of 
stars in And~I and And~III exists \citep{guh00}, and two such studies of 
And~II exist, both being based on the same data set \citep{cote99a,cote99b}.  For And~I 
and III, \cite{guh00} were able to measure velocities for 29 and 7 dSph member 
stars, respectively, but with large velocity errors.  \cite{cote99a} established 
accurate radial velocities for seven member stars in And II and confirmed 35 
other members with very poor radial velocity accuracy, $\sigma_v >$ 40 km~s$^{-1}$ 
(Cote et~al.\ 1999b; P. Cote 2006, private communication).  The first aim of our 
survey is to significantly increase the numbers of stars with accurate radial 
velocities ($\sigma_v \lesssim$ 10~km~s$^{-1}$) in these galaxies.  In 
Figure~\ref{fig:giant} (bottom), we present radial velocity histograms for all 
objects for which we could measure a velocity as discussed earlier (grey histogram).  
For each of our And I, II, and III fields, there is a very clean detection of a kinematically 
cold spike representing stars that belong to the dSph.  We also see a small population 
of field M31 stars with a broad range of velocities.  Considering that each of 
And I, II, and III are located well beyond the radius at which M31's 
halo dominates its inner spheroid (25 -- 30~kpc, Guhathakurta et~al.\ 2005; Kalirai et~al.\ 
2006a), these field stars likely represent M31 halo members.  Finally, there is a 
population of foreground Milky Way dwarf stars at small negative velocities.  

We establish membership of stars belonging to And~I, II, and III using a three step process.  
This involves first isolating the sample of stars that have characteristics of 
giants at the distance of M31, second, eliminating any M31 substructure, and, third, 
cleaning the dSph populations of any M31 field stars in the halo.  For the first 
cut, we can use photometric and spectroscopic diagnostics to remove any Milky Way 
dwarf stars along the line of sight.  The diagnostics used for this separation 
include radial velocity measurements, $DDO51$ photometry, strength of the Na\,{\sc i} 
doublet, position of the star in the CMD, and a comparison of the photometric and 
spectroscopic metallicities of each star.  Details on the resolving power of each of 
these diagnostics to weed out Milky Way dwarf star contamination, as well as a 
careful step-by-step cookbook on how they are applied to the data set, are presented 
in \cite{gilbert06}.  As those methods were developed to separate out stars in M31's 
(kinematically hot) field halo from Milky Way contaminants, we note that we made two small 
changes in this application.  First, the velocity and dispersion of the expected dSph 
stars were set to the rough values of the peaks in Figure~\ref{fig:giant}, and second 
the distance of each satellite (e.g., in calculating a photometric metallicity) was 
set to the known distances of And~I (745 $\pm$ 24~kpc), And~II (652 $\pm$ 18~kpc), and 
And~III (749 $\pm$ 24~kpc) as reported in \cite{mcconnachie05}.  The uncertainty 
in the adopted distance of each satellite produces a very small effect on the overall 
classifications, given both the nature of the measurement and the relative insensitivity of 
the two affected diagnostics (i.e., metallicity comparison and position in the CMD) compared 
to others (e.g., radial velocity and $DDO51$ photometry).  The final likelihoods 
of a star being a giant or dwarf are illustrated in the top panel of Figure~\ref{fig:giant}; 
values of $L_i$ that are positive are more likely to be giants and values with $L_i$ $<$ 0 
are likely Milky Way dwarfs.  Our final selection of giants are those stars that are three 
times more likely to be giants than Milky Way dwarfs (corresponding to $L_i$ $\geq$ 0.5), as 
discussed in \cite{gilbert06} and \cite{kalirai06a}.

The distribution in radial velocity of giant stars in fields And~I, II, and III are shown as the 
solid histogram in Figure~\ref{fig:giant} (bottom).  Clearly, this separation has removed the 
obvious Milky Way dwarf stars at small negative velocities, however, we note that a 
few dwarf outliers are also removed at (or close to) the velocity of the dSph satellites.  
The distribution of the M31 giants (filled points) and Milky Way dwarfs (open points) are also shown on 
the CMD in Figure~\ref{fig:giant} (middle panels).  The giants form a tight sequence in 
the CMD, stretching from the tip of the RGB to our photometric limit at $I_{\rm 0} \sim$ 
22, whereas the dwarf stars occupy a broad range in color and luminosity depending on 
their spectral type and distance.  The three solid red curves on these diagrams are 
theoretical isochrones for an age of 12~Gyr and [$\alpha$/Fe] = 0.0, with a metallicity 
of [Fe/H] = $-$2.31 (left), $-$1.31, and $-$0.83, and $-$0.40 (right) from the 
\cite{vandenberg06} models.  The dashed curves are the same isochrones assuming 
[$\alpha$/Fe] = +0.3. As we will quantify in \S\,\ref{abundancedistributions}, each of these dSphs 
appear to contain a metal-poor population of stars with a modest dispersion.


\begin{figure}
\begin{center}
\leavevmode
\includegraphics[width=8.7cm]{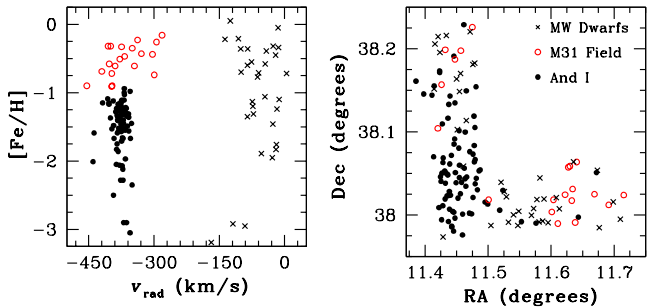}
\end{center}
\caption{The And~I dSph is spatially located within identified halo substructure 
from the Giant Southern Stream \citep{ibata07,gilbert09}.  {\it Left} -- The 
substructure is isolated from the dSph stars with a metallicity cut, where the 
more metal-rich material is associated with the stream debris.  
{\it Right} -- As expected, these stars, shown as red open circles, are also located 
in the outskirts of our spectroscopic masks which intersect at the center of And~I.}
\label{fig:member2d1}
\end{figure}



\begin{figure*}
\begin{center}
\leavevmode
\includegraphics[width=10cm,angle=270, bb = 22 16 463 784]{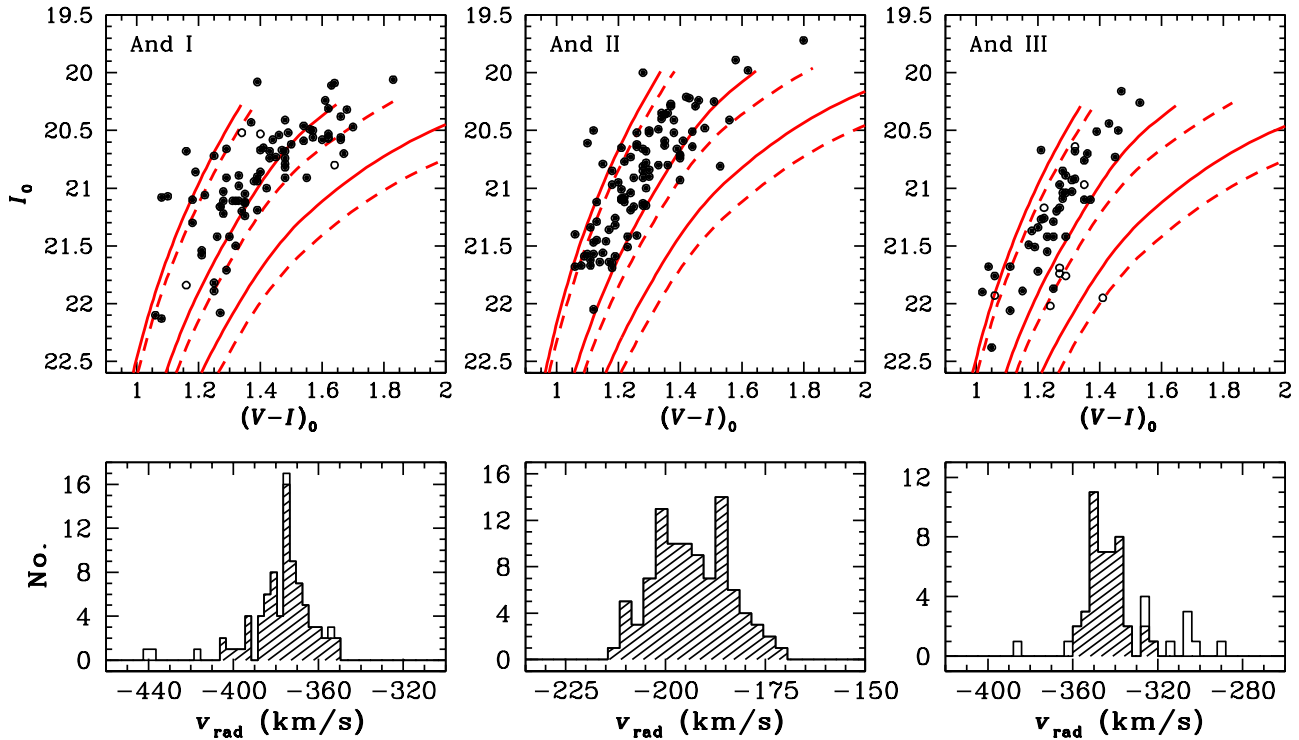}
\end{center}
\caption{A closer view of the CMD ({\it top}) and radial velocity histograms ({\it bottom}) for 
our final And~I (80 stars), And~II (95 stars), and And~III (43 stars) samples of RGB 
members (filled points and histograms).  Open points and histograms show the M31 
field halo giant stars that are eliminated from the final sample as discussed in 
\S\,\ref{membership}.  The three sets of isochrones shown in each panel of the 
CMDs are taken from \cite{vandenberg06}, for an age of 12~Gyr and metallicities of 
[Fe/H] = $-$2.31 (left), $-$1.31 (middle), and $-$0.83 (right), and have been 
shifted to each satellites' distance modulus.  The models for [$\alpha$/Fe] = 
0.0 are shown as solid curves and those for [$\alpha$/Fe] = +0.3 are shown as 
dashed curves.  Clearly, all three of these satellites are metal-poor systems.}
\label{fig:cmdvel}
\end{figure*}


The second cut in our membership criteria only applies to And~I, which is situated at 
a spatial location in M31 that appears to contain substructure related to the Giant 
Southern Stream \citep{ibata07}.  As \cite{gilbert09} show, the substructure 
in this field is located at roughly the same velocity as And~I, but is much more 
metal-rich than the dSph stars.  We can see the substructure in the CMD of this dSph 
as an extention of metal-rich stars.  This is illustrated more clearly in Figure~\ref{fig:member2d1}, 
where we show the M31 giants (circles) and Milky Way dwarfs (crosses) in the And~I field on an [Fe/H] 
vs $v_{\rm rad}$ plane (left panel -- see \S\,\ref{abundancedistributions} for the details 
on how [Fe/H] is calculated).  The reported photometric metallicity of 
the Milky Way dwarfs is meaningless since their distance is unknown.  However, the metallicity of 
And~I and the stream contamination is correct since these populations are both located 
at M31's distance.  The diagram shows a bi-modality with a split at 
[Fe/H] $\sim$ $-$0.9, where the bulk of the And~I stars (filled points -- see below) comprise 
the more metal-poor, tight grouping of stars.  To further illustrate that the metal-rich stars 
in this diagram are likely not members of And~I, we look at their positions relative to our 
two spectroscopic masks in the right panel.  As expected, the stars associated with the 
stream contamination are found in the outskirts of the masks, far from And~I, where 
the densities of the dSph and the field become more equal.  We note that the King limiting 
radius of the dSph is 10$'$, and most of these stars are located beyond this radius.  Formally, 
we have selected the exact cut that separates the filled circles from the open circles 
arbitrarily at the apparent break point in Figure~\ref{fig:member2d1} (left).  For our sample, 
this translates to a hard cut of [Fe/H] $<$ $-$0.92 for the And~I members, although we recognize 
that the exact separation may be slightly shifted from this value and depends on the relative 
age and distance difference between the two populations.  The hard cut may also 
lead to the truncation of any metal-rich tail in the dSphs metallicity distribution function.

The final cut in our membership criteria eliminates any likely M31 field halo giants not 
associated with each dSph.  The limiting radii of And~I is 10.4$'$, of And~II is 22.0$'$, 
and of And~III is 7.2$'$ \citep{mcconnachie06}, and so we eliminate three stars in And~I 
and eight stars in And~III that are located beyond these limits.  These eliminations 
work to actually remove stars that can be easily identified as outliers on the 
velocity histograms.  For example, in And~I one of the stars eliminated has a radial 
velocity of $-$440.0~km~s$^{-1}$, and in And~III one of the stars is moving at 
$v_{\rm rad}$ = $-$431.6~km~s$^{-1}$.  Both of these stars have velocities 
that are 5 -- 10$\sigma$ deviant from the final observed dispersion.  Additionally, 
the cut at the limiting radius in And~III eliminates the two giants with $V{\rm-}I$ $>$ 2 
in the CMD in Figure~\ref{fig:giant}.  The photometric metallicity of 
these two stars is calculated 
to be [Fe/H] $>$ $-$0.25, well over 10 times more metal-rich than the remaining dSph 
stars (see \S\,\ref{photmet}).  The resulting velocity distribution still includes M31 
field halo giants that are spatially superimposed on the dSph population. To eliminate these 
last few stars, we measure the mean and sigma of the resulting velocity distribution 
and iteratively eliminate stars that are more than 3-$\sigma$ from the mean.  This 
iteration removes just two stars from And~I, zero stars from And~II, and 
three stars from And~III's sightline.  The eliminated stars were again not centrally 
located in the dSph suggesting they are likely field halo giants.  The final selected 
members of these three dSphs include 80 stars in And~I, 95 stars in And~II, and 43 
stars in And~III.  

We stress that our primary goal from the selection process described above 
is to construct as secure a sample as possible of confirmed RGB members of these three 
satellites.  Given the cuts used, it is possible (although unlikely) that we 
have in fact thrown a member or two out of the sample, however our sample sizes 
are large enough that the exclusion of a few stars will not impact the analysis 
that follows.  The filled histograms in Figure~\ref{fig:cmdvel} (bottom) illustrate 
the velocities for the stars in the final sample, and the filled points in the top panel 
of this Figure show their distribution in the CMD.  The open histograms (bottom) and 
open points (top) in Figure~\ref{fig:cmdvel} represent the eliminated giants from 
the limiting radius and $\sigma$-clipping cuts discussed above.

\section{Mean Velocities, Intrinsic Velocity Dispersions, and $M/L$ Ratios} \label{meanvelocity}

We calculate the mean radial velocity and intrinsic dispersion for each of the six 
dSphs using the maximum-likelihood method described by \cite{walker06}.  This 
method assumes that the observed dispersion is the sum of the true intrinsic 
dispersion and the velocity errors discussed at length in \S\,\ref{RVuncertainties}.  
For And~I, we find $v_{\rm rad}$ = $-$375.8 $\pm$ 1.4~km~s$^{-1}$ and $\sigma_v$ = 10.6 $\pm$ 1.1~km~s$^{-1}$, 
for And~II we find $v_{\rm rad}$ = $-$193.6 $\pm$ 1.0~km~s$^{-1}$ and $\sigma_v$ = 7.3 $\pm$ 0.8~km~s$^{-1}$, 
for And~III we find $v_{\rm rad}$ = $-$345.6 $\pm$ 1.8~km~s$^{-1}$ and $\sigma_v$ = 4.7 $\pm$ 1.8~km~s$^{-1}$, 
for And~VII we find $v_{\rm rad}$ = $-$309.4 $\pm$ 2.3~km~s$^{-1}$ and $\sigma_{v}$ = 9.7 $\pm$ 1.6~km~s$^{-1}$,
for And~X we find $v_{\rm rad}$ = $-$168.3 $\pm$ 1.2~km~s$^{-1}$ and $\sigma_{v}$ = 3.9 $\pm$ 1.2~km~s$^{-1}$,
and for And~XIV we find $v_{\rm rad}$ = $-$481.0 $\pm$ 1.2~km~s$^{-1}$ and $\sigma_{v}$ = 5.4 $\pm$ 1.3~km~s$^{-1}$.  
The mean radial velocities of these six galaxies are consistent with M31 membership, since M31 has 
a systemic velocity of $-$300~km~s$^{-1}$.  The one exception to this may be And~XIV, which has both a large 
negative radial velocity and a large distance from M31 (both projected and along the line of sight).  As 
discussed in \cite{majewski07},  the galaxy may therefore be falling into the Local Group for the first 
time.  The velocities for And~I and II are inconsistent with the measurements by previous studies within the 
mutual 1-$\sigma$ errorbars established by this and the previous studies \citep{guh00,cote99b}.  As we 
noted earlier, these previous results are only based on a handful of stars and therefore our values 
are much more secure.  For And~III, \cite{guh00} found $v_{\rm rad}$ = $-$352.3 $\pm$ 
13.6~km~s$^{-1}$ based on 9 RGB members, which is consistent with our much more 
precise result.

The central velocity dispersion of the six galaxies in this study varies by more 
than a factor of 2.5.  The only previous measurement to compare our results with 
is the high resolution study of \cite{cote99b}, which established $\sigma_v$ = 
9.3 $\pm$ 2.7~km~s$^{-1}$ for And~II based on seven RGB stars.  Our result is 
consistent within the larger error bar determined by \cite{cote99b} for their 
measurement, however we find that the velocity dispersion is lower.  We can use 
our new velocity dispersions to put first order constraints on the total mass 
(stellar and dark matter) of each satellite using the method described by \cite{illingworth76}, 
$M$ = 167*$\beta$$r_c$$\sigma_v^2$ where $\beta$ = 8 for dSphs \citep{mateo98}.  
This formalism was initially constructed for the analysis of line of sight velocities 
in globular clusters and therefore has several assumptions built into it, all of 
which may not be true.  For example, we are assuming the galaxies are spherical, 
are in dynamical equilibrium, and have an isotropic velocity dispersion.  We are 
also assuming that the stellar distribution follows a King profile (which is 
observed) and traces the dark matter.  We return to a discussion on how more accurate 
dynamical mass measurements can be established for M31 dSphs below.

The core radius of four of the six dSphs is known from the structural study of 
\cite{mcconnachie06}, who find $r_{c}$ = 580 $\pm$ 60~pc for And~I, $r_{c}$ = 990 $\pm$ 40~pc 
for And~II, $r_{c}$ = 290 $\pm$ 40~pc for And~III, and $r_{c}$ = 450 $\pm$ 20~pc for And~VII 
(geometric means).  The measured core radius, and an assumed 10\% uncertainty, of And~X is 
$r_c$ = 270 $\pm$ 30~pc (Zucker et~al.\ 2007) and that of And~XIV is much larger at 
$r_c$ = 730 $\pm$ 70~pc (Majewski et~al.\ 2007).  The total masses of these satellites are therefore 
$M$ = (8.7 $\pm$ 1.6) $\times$ 10$^{7}$ $M_\odot$ for And~I, 
$M$ = (7.0 $\pm$ 1.1) $\times$ 10$^{7}$ $M_\odot$ for And~II, 
$M$ = (8.6 $\pm$ 4.8) $\times$ 10$^{6}$ $M_\odot$ for And~III, 
$M$ = (5.7 $\pm$ 1.3) $\times$ 10$^{7}$ $M_\odot$ for And~VII, 
$M$ = (5.5 $\pm$ 2.5) $\times$ 10$^{6}$ $M_\odot$ for And~X, and 
$M$ = (2.9 $\pm$ 1.0) $\times$ 10$^{7}$ $M_\odot$ for And~XIV.  
At the luminous end, for $M_{\rm V}$ = $-$11 to $-$13, the masses of And~I, II, and VII are 
similar to Milky Way dSphs (e.g., Fornax and Leo~I), calculated in the same way.  
At lower luminosities, the masses of And~III and X appear to be lower than systems in 
the Milky Way with similar brightness, such as Carina, Draco, and Sextans. 

We can calculate the mass-to-light ratios in Solar units of And~I, II, III, VII, X, and 
XIV by combining the masses determined above with the measured luminosities of each 
satellite (see McConnachie \& Irwin 2006; Zucker et~al.\ 2007; Majewski et~al.\ 2007).  
This gives 
$M/L_{\rm V}$ = 19 $\pm$ 4 for And~I, 
$M/L_{\rm V}$ = 7.5 $\pm$ 1.8 for And~II, 
$M/L_{\rm V}$ = 8.3 $\pm$ 5.2 for And~III, 
$M/L_{\rm V}$ = 3.2 $\pm$ 1.2 And~VII, 
$M/L_{\rm V}$ = 37 $\pm$ 24 for And~X, and 
$M/L_{\rm V}$ = 160 $\pm$ 95 for And~XIV.  
The mass-to-light ratios indicate that these satellites are dark matter dominated, as expected, 
however both And~II and III are at the low end of the range of $M/L$ ratios for dSphs.   A 
more in depth comparison of the results in this section, as well as their implications, will be 
presented in \S\,\ref{DMmasses}.

In Table~2 we summarize several of the fundamental properties of each of the six satellites 
discussed above.  This includes their luminosities, projected distance from M31, mean radial 
velocities, intrinsic velocity dispersions, metallicities, and metallicity dispersions (see 
\S\,\ref{chemabund} for half-light radii, total masses, and $M$/$L$ ratios).  Further details 
on the individual galaxies are also available in the \cite{guh00}, \cite{kalirai09}, and 
\cite{majewski07} studies.



\begin{table*}
\begin{center}
\caption{}
\begin{tabular}{lccccccccc}
\hline
\hline
\multicolumn{1}{l}{Property} & \multicolumn{1}{c}{And~I} & \multicolumn{1}{c}{And~II} & \multicolumn{1}{c}{And~III} &
\multicolumn{1}{c}{And~VII} & \multicolumn{1}{c}{And~X} &\multicolumn{1}{c}{And~XIV} \\
\hline
$M_{\rm V}$                      &            $-$11.8 $\pm$ 0.1 & $-$12.6 $\pm$ 0.2 & $-$10.2 $\pm$ 0.3 & $-$13.3 $\pm$ 0.3 & $-$8.1 $\pm$ 0.5 &  $-$8.3 $\pm$ 0.5 \\                                                                                                   
$R_{\rm proj}$ (kpc)              &           44.9 & 145.6  & 68.2   & 214.5  & 75.5   & 162.5 \\                                                                                                                                                                       
$v_{\rm rad}$ (km~s$^{-1}$)       &           $-$375.8 $\pm$ 1.4 & $-$193.6 $\pm$ 1.0 & $-$345.6 $\pm$ 1.8 & $-$309.4 $\pm$ 2.3 & $-$163.8 $\pm$ 1.2 & $-$481.0 $\pm$ 2.0 \\                                                                                            
$\sigma_v$ (km~s$^{-1}$)         &            10.6 $\pm$ 1.1   &  7.3 $\pm$ 0.8  &  4.7 $\pm$ 1.8  &  9.7 $\pm$ 1.6  &  3.9 $\pm$ 1.2  &  5.4 $\pm$ 1.3 \\

[Fe/H]  &       $-$1.45 $\pm$ 0.04 & $-$1.64 $\pm$ 0.04 & $-$1.78 $\pm$ 0.04 & $-$1.40 $\pm$ 0.30 & $-$1.93 $\pm$ 0.11 & $-$2.26 $\pm$ 0.05 \\                                                                                            
$\sigma_{[Fe/H]}$                &            0.37 & 0.34 & 0.27 & \nodata & 0.48 & \nodata  \\                                                                                                                                                                         
$r_{c}$\tablenotemark{1} & 580 $\pm$ 60 & 990 $\pm$ 40 & 290 $\pm$ 40 & 450 $\pm$ 20 & 271 $\pm$ 27 & 734 $\pm$ 73 \\
2D $R_{e}$ (pc)\tablenotemark{2} &          682 $\pm$ 57  & 1248 $\pm$ 40 & 482 $\pm$  58 & 791  $\pm$ 45 & 339 $\pm$ 6 & 413 $\pm$ 41 \\                                                                                                                            
3D $r_{1/2}$ (pc)\tablenotemark{3}   &          900 $\pm$ 75  & 1659 $\pm$ 53 & 638 $\pm$  77 & 1050 $\pm$ 60 & 448 $\pm$ 8 & 461 $\pm$ 155 \\                                                                                                                            
$M$ ($M_\odot$)\tablenotemark{4} &            (8.7 $\pm$ 1.6) $\times$ 10$^{7}$ & (7.0 $\pm$ 1.1) $\times$ 10$^{7}$ & (8.6 $\pm$ 4.8) $\times$ 10$^{6}$ & (5.7 $\pm$ 1.3) $\times$ 10$^{7}$ & 
(5.5 $\pm$ 2.5) $\times$ 10$^{6}$ & (2.9 $\pm$ 1.0) $\times$ 10$^{7}$ \\  
$M$/$L$ ($M_\odot$/$L_\odot$)\tablenotemark{4}  &       19  $\pm$ 4 & 7.5 $\pm$ 1.8 & 8.3 $\pm$ 5.2 & 3.2 $\pm$ 1.2 & 37  $\pm$  24 & 160 $\pm$ 95  \\                                                                                                                            
$M_{1/2}$ ($M_\odot$)\tablenotemark{5}          &     (7.0 $\pm$ 1.2) $\times$ 10$^{7}$ & (6.1 $\pm$ 1.0) $\times$ 10$^{7}$ & (9.6 $\pm$ 5.4) $\times$ 10$^{6}$ & (6.9 $\pm$ 1.6) $\times$ 10$^{7}$ & 
(4.7 $\pm$ 2.0) $\times$ 10$^{6}$ & (9.2 $\pm$ 4.4) $\times$ 10$^{6}$ \\
$M_{1/2}$/$L_{1/2}$ ($M_\odot$/$L_\odot$)\tablenotemark{5} & 31 $\pm$  6 &  13 $\pm$  3 &  19 $\pm$ 12 &  7.7 $\pm$ 2.8 &  63  $\pm$ 40 &  102 $\pm$ 71 \\
No. Stars                                      &                80 & 95 & 43 & 18 & 22 & 38 \\                                                                                                                                                                                            
\hline
\end{tabular}
\tablenotetext{1}{$r_{c}$ is the core radius, the uncertainty is taken to be 10\% for And~X and And~XIV.}
\tablenotetext{2}{$R_{e}$ is the two dimensional elliptical half light radius.}
\tablenotetext{3}{$r_{1/2}$ is the three dimensional deprojected half light radius.}
\tablenotetext{4}{These masses represent the total system mass, as measured using the \cite{illingworth76} formalism.}  
\tablenotetext{5}{The masses at the half light radius are calculated using $M_{1/2}$ = 3$G^{-1}$$\langle$$\sigma_v^2$$\rangle$$r_{1/2}$ \citep{wolf09}.}
\label{table:params2}
\end{center}
\end{table*}

\subsection{More Accurate Masses and Mass \\ Profiles of M31's dSphs} \label{accuratemasses}

Both \cite{wolf09} and \cite{walker09a} have recently compared the masses of Milky Way dSphs 
calculated using the method above to those derived using new mass estimators for 
dispersion-supported systems that result from a manipulation of the Jeans equation.  
The new mass estimators are shown to place tight constraints 
on the system mass within the half light radius, and weaker constraints at larger 
radii.  In the \cite{walker09a} analysis, the mass at the half light radius 
using the new formalism is found to be less than the \cite{illingworth76} (total 
mass) approximation by $\sim$50\% (systematically).  \cite{wolf09} go one step 
further and compare the estimated total masses of dSphs (see below) to the 
\cite{illingworth76} relation, and find that the latter underpredicts both 
the total mass of dSphs and the associated uncertainties in the mass.  The reason 
for this is related to the assumed distribution of the total mass, where the 
\cite{illingworth76} approximation forces the mass distribution to truncate near 
the stellar extent of the galaxy.  However, over a range of almost two orders of 
magnitude in mass, the offset in mass is nearly constant (see their Figure~C1 in the 
Appendix).  

The baseline for the comparisons above that demonstrates the mass at the half light 
radius is well constrained in these new derivations comes from the full modeling 
of {\it individual} radial velocities to yield mass profiles for Milky Way dSphs.  For 
example, \cite{strigari08b} assume the radial velocities are related to the overall mass 
distribution through the Jeans equation, and that the dark matter follows a five 
parameter density profile.  They also allow the velocity anisotropy to vary in the 
modeling, however, the systems are still assumed to be spherically symmetric and in 
dynamical equilibrium (see also Walker et~al.\ 2009a).  Given this recent work, we 
calculate the mass at the half light radius for each of the M31's dSphs using the 
new formalism presented in \cite{wolf09}, $M_{1/2}$ = 
3$G^{-1}$$\langle$$\sigma_v^2$$\rangle$$r_{1/2}$, where $r_{1/2}$ is the three dimensional 
deprojected half light radius (see Table~2).  This calculation yields
$M_{1/2}$ = (7.0 $\pm$ 1.2) $\times$ 10$^{7}$ $M_\odot$ for And~I, 
$M_{1/2}$ = (6.1 $\pm$ 1.0) $\times$ 10$^{7}$ $M_\odot$ for And~II, 
$M_{1/2}$ = (9.6 $\pm$ 5.4) $\times$ 10$^{6}$ $M_\odot$ for And~III, 
$M_{1/2}$ = (6.9 $\pm$ 1.6) $\times$ 10$^{7}$ $M_\odot$ for And~VII, 
$M_{1/2}$ = (4.7 $\pm$ 2.0) $\times$ 10$^{6}$ $M_\odot$ for And~X, and 
$M_{1/2}$ = (9.2 $\pm$ 4.4) $\times$ 10$^{6}$ $M_\odot$ for And~XIV.
These results are summarized in Table~2 and discussed further in \S\,\ref{DMmasses}.  
Additional analysis of the mass profiles of these satellites in comparison to 
Milky Way dSphs will also be presented in Wolf et~al.\ (2010, in prep.), 


\section{Chemical Abundances} \label{chemabund}

Da Costa et~al.\ (1996; 2000; 2002) measured the abundances of And~I, II, and III 
from {\it HST}/WFPC2 photometric data and found that all three dSphs are metal-poor, 
([Fe/H] $<$ $-$1.45), but with large variations in the internal abundance spread.  For 
And~I, they found a very high dispersion of $\sigma_{\rm [Fe/H]}$ = 0.60, whereas for 
And~III, they find that $\sigma_{\rm [Fe/H]}$ = 0.12.  For the more luminous satellite 
And~II, Da~Costa et~al.\ find $\sigma_{\rm [Fe/H]}$ = 0.36.  These large variations 
suggest that the M31 dSphs may have experienced quite different evolutionary histories.  
For example, a galaxy such as And~III would have had a tough time retaining its 
enrichment products whereas a system such as And~I may have experienced either 
multiple epochs (or an extended epoch) of star formation.  Da Costa et~al.\ stress 
the need for an independent spectroscopic study of RGB stars in these satellites.  
As we've already demonstrated in \S\,\ref{membership}, the And~I sightline contains bright 
red giants that belong to the debris of the Giant Southern Stream.  These stars are 
much more metal-rich than the dSph population, and if not removed, will artificially 
inflate the measured abundance spread.

As in our previous studies of M31's stellar halo (e.g., Kalirai et~al.\ 2006a), we 
determine the chemical abundance and abundance spread of stars in And~I, II, and 
III using two independent methods.  Both methods are used on the restricted sample 
of confirmed RGB member stars in each satellite, as determined in \S\,\ref{membership}.  
The first method is based on a comparison of the location of the dSph stars on the 
($I$, $V{\rm-}I$) CMD to a new set of theoretical isochrones.  The second method is 
based on a spectroscopic measurement of the equivalent width of the Ca\,{\sc ii} 
triplet ($\lambda$ $\sim$ 8500~${\rm \AA}$).  The advantages of this two-pronged photometric 
{\it and} spectroscopic approach is that (1) a comparison of two independent 
measures can shed light on the existence of a systematic bias in either method, 
(2) the radial velocity measurements from the spectroscopic data can be used to 
ensure that the photometric and spectroscopic abundance distribution consists 
only of true dSph RGB stars (i.e., they are uncontaminated by Milky Way dwarfs), 
and (3) the spectra for like stars can be coadded to yield information on detailed 
chemical abundances (e.g., [$\alpha$/Fe]).  Below, we discuss each of these two 
methods in turn to derive the abundances of stars in And~I, II, and III.


\subsection{Photometric Metallicity Determination}\label{photmet}

The positions of stars along the RGB in a stellar population can be used to 
determine their chemical abundances, provided their age is known.  In an 
($I$, $V{\rm-}I$) CMD, the shape of the RGB is such that metal-rich stars 
become increasingly redder relative to metal-poor stars as their luminosity 
increases.  The stars that we targeted in And~I, II, and III are among the 
brightest RGB stars in these dSphs, and therefore we take full advantage of 
this color sensitivity.

The spectroscopically selected final sample of member stars in And~I, II, 
and III are shown as filled circles in the top panel of Figure~\ref{fig:cmdvel}.  
As discussed above, the red curves represent theoretical stellar isochrones from 
the new models of \cite{vandenberg06}, for three different metallicities, 
[Fe/H] = $-$2.31 (left), $-$1.31 (middle), and $-$0.83 (right).  The solid curves 
are models with no alpha-enhancement and the dashed curves assume [$\alpha$/Fe] = 
$+$0.3.  The models have been shifted to the known distance to each satellite, as 
measured from the tip of the red giant branch.  
We measure the metallicity distribution function for each satellite by interpolating 
the magnitude and color of each member star within a larger grid of similar 
isochrones, including a dozen models within the range [Fe/H] = $-$2.31 to $-$0.83.  
For the few stars in each satellite that are located outside the bounds of these 
models, we extrapolate the metallicity measurements.  The errors on these 
measurements therefore depend both on the distance uncertainty to the dSph 
and the error in the photometry.

The photometric metallicity measurements of And~I, II, and III assume a particular 
set of models, in this case the new \cite{vandenberg06} isochrones.  We can also 
recalculate the metallicities assuming a different isochrone set.  For example, 
our results agree with the metallicities that would be derived from either the 
Padova models \citep{girardi02} or the Yale-Yonsei models (Y$^2$;  Demarque 
et~al.\ 2004) at the 0.15 dex level.  The metallicities also depend on the 
assumed age of the stellar population.  For intermediate to old ages, there is 
little color sensitivity of an RGB star with age.  For example, a shift in the entire age of the 
dSph from 12~Gyr (our adopted age) to 6~Gyr would translate to a $+$0.2~dex offset in 
[Fe/H]$_{\rm phot}$.  Although there currently exist no direct age measurements 
for And~I, II, or III from resolved main-sequence turnoff fitting, several hints 
suggest that the bulk of the stellar populations in each of these three dSphs 
are old.  For example, the CMDs in Da~Costa et~al.\ are similar to those of Galactic globular 
clusters, which are all old.  Specifically, the galaxies show an absence of upper 
asymptotic giant branch stars, contain blue horizontal branch stars, and contain 
RR Lyrae stars.

\subsection{Spectroscopic Metallicity Determination}\label{specmet}

The spectroscopic metallicity of a star, which we designate [Fe/H]$_{\rm spec}$, 
can be derived from the equivalent widths of the three Ca\,{\sc ii} absorption 
lines.  The independent measurements of the three lines are combined to produce 
a reduced equivalent width as described in Rutledge, Hesser, \& Stetson (1997).  
This reduced equivalent width is then empirically calibrated based on 
observations of Galactic globular cluster RGB stars to yield [Fe/H]$_{\rm spec}$ 
\citep{rutledge97b}.  The correction for surface gravity is 
derived from the luminosity of the HB in the {\it HST\/} imaging 
study of each dSph, which we take from the Da Costa et~al.\ work.  We 
measure this from their CMDs to be $V_{\rm HB}$ = 25.20 for And~I, 
$V_{\rm HB}$ = 24.85 for And~II, and $V_{\rm HB}$ = 24.95 for And~III.

\begin{figure}
\begin{center}
\leavevmode
\includegraphics[width=8.5cm]{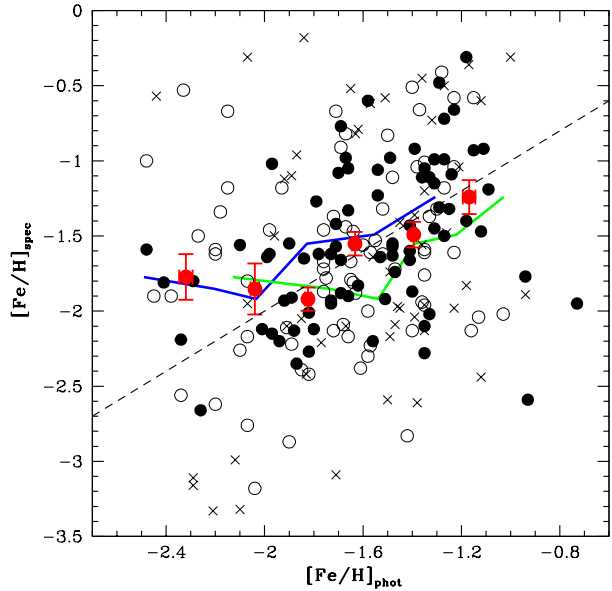}
\end{center}
\caption{The [Fe/H]$_{\rm phot}$ measurements are compared to the independently 
measured [Fe/H]$_{\rm spec}$ values, over a 1.5~dex spread in [Fe/H]$_{\rm phot}$.  
The dashed line illustrates equality between the two measurements, and, in general, our 
data points follow this trend nicely for [Fe/H]$_{\rm phot}$ $>$ $-$2.2.  The scatter 
among the spectroscopically measured abundances is large, as expected given the low 
$S/N$ of our spectra.  This is illustrated by plotting the symbols as a function of 
the spectral $S/N$, where crosses represent all stars with a spectral $S/N$ $<$ 3, 
open circles represent stars with 3 $<$ $S/N$ $<$ 5, and filled circles represent 
stars with $S/N$ $>$ 5.  The scatter in [Fe/H]$_{\rm spec}$ is clearly reduced among 
the better quality data.  The larger data points with error bars are binned averages 
of the $S/N$ $>$ 3 data, where the photometric metallicities have been calculated using 
the \cite{vandenberg06} isochrones with an age of 12~Gyr and [$\alpha$/Fe] = 0.0.  
The two solid blue and green lines show the same comparison for different assumptions 
on age and [$\alpha$/Fe], as discussed in \S\,\ref{specmet}.} \label{fig:metcomp}
\end{figure}

The uncertainty in [Fe/H]$_{\rm spec}$ from our low $S/N$ spectra results in 
a larger scatter than the [Fe/H]$_{\rm phot}$ measurements.  Still, we can 
use this measurement to provide an independent check on our overall 
metallicity measurements.  In Figure~\ref{fig:metcomp}, we 
compare the two metallicity measurements to one another.  We have shown 
all individual data points, measured using both techniques, as well as 
binned averages of the spectroscopic measurements (larger filled circles with 
error bars).  These were calculated using a minimum bin size of 0.2 dex in 
[Fe/H]$_{\rm phot}$ while ensuring $>$15 stars in each bin, and requiring the 
spectra of the individual stars to have at least $S/N$ $>$ 3 (see below).  Out 
of our initial sample of 218 member stars in And~I, II, and III, this $S/N$ cut 
leaves 158 objects.  We reduce this sample by three more stars to eliminate 
objects for which the [Fe/H]$_{\rm spec}$ values were unrealistically high 
($+$0.71, $+$2.05, and $+$2.61).  There is a nice overall agreement between the 
resulting mean values of the two measurements over a range that includes most 
of the data points, [Fe/H]$_{\rm phot}$ = $-$2.2 to $-$1.1.  The dashed line 
illustrates the 1:1 relation.  For our most metal-poor bin, [Fe/H]$_{\rm phot}$ 
$<$ $-$2.2, the spectroscopic metallicities are systematically more metal-rich 
than the photometric values.  For stars with these low metallicities, the 
Ca\,{\sc ii} lines are very shallow and the spectroscopic metallicities using 
this method are likely in error.  In fact, Kirby et~al.\ (2008) show that the 
spectroscopic metallicities of similar stars in Galactic dSphs have been 
overestimated in past studies.


We also illustrate how the independent photometric and spectroscopic metallicity 
measurements compare for various $S/N$ cuts.  The crosses in Figure~\ref{fig:metcomp} 
illustrate those objects with $S/N$ $<$ 3, the open circles represent stars with 3 $<$ 
$S/N$ $<$ 5, and the filled points are those with $S/N$ $>$ 5.  The higher $S/N$ points 
tend to be clustered closer to the 1:1 line as compared to the measurements 
from poorer quality spectra, indicating that much of the vertical scatter 
in the diagram results from poor characterization of the Ca\,{\sc ii} triplet 
lines in these spectra.  


\begin{figure*}
\begin{center}
\leavevmode
\includegraphics[width=10cm,angle=270, bb = 22 16 446 784]{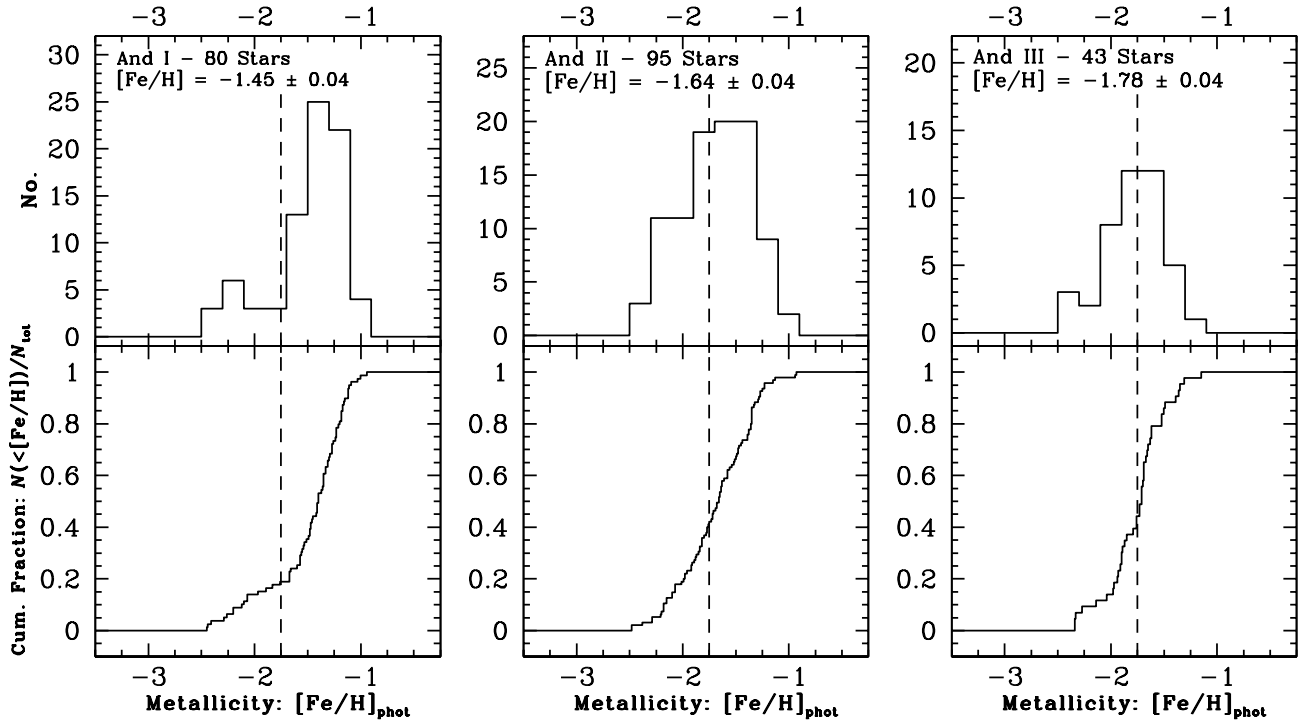}
\end{center}
\caption{The discrete photometric metallicity distribution functions (MDF) for And~I, II, 
and III ({\it top}) and the corresponding cumulative distributions ({\it bottom}), based 
on our confirmed sample of member stars in each dSph.  The dashed line is held fixed at 
[Fe/H]$_{\rm phot}$ = $-$1.75 in each panel as a guide.  We find that And~III is the most metal-poor 
dSph of the three, and that And~I is the most metal-rich.  The metallicity dispersion in 
all three dSphs is similar, $\sigma_{\rm [Fe/H]}$ = 0.3 -- 0.4.}
\label{fig:abunddist}
\end{figure*}


To summarize, we find a nice overall agreement between our photometric and 
spectroscopic metallicity measurements for these satellites, indicating that 
any systematic biasses are likely small.  For example, if we assume the dSphs 
are significantly younger than 12~Gyr, the resulting comparison of metallicities 
leads to a relation offset than the 1:1 line.  We illustrate this, for an age 
of 6~Gyr, as the green curve in Figure~\ref{fig:metcomp}.  We also investigate 
the effects of different $\alpha$-enhancements.  The photometric metallicities 
shown in Figure~\ref{fig:metcomp} assume these three dSphs are not enhanced in 
$\alpha$-elements relative to the Sun (e.g., we used the isochrones with 
[$\alpha$/Fe] = 0.0 for these calculations).  Stars in the Milky Way's field halo, 
and globular clusters, are known to be $\alpha$-enhanced ([$\alpha$/Fe] = $+$0.3) 
and it is generally believed that this is a result of the stars in the halos of 
galaxies forming in early ``bursts''.  Milky Way dSphs also show some 
$\alpha$-enhancement, although not as high as that measured in globular clusters.  For 
example, \cite{shetrone01} find $+$0.02 $<$ [$\alpha$/Fe] $<$ $+$0.13 for the 
three dSphs Draco, Sextans, and Ursa Minor.  A mild $\alpha$-enhancement, such 
as that seen in these Milky Way galaxies, produces a very small affect on our 
metallicity measurements.  However, we note that if And~I, II, and III are 
enhanced in $\alpha$-elements at a level similar to Galactic globular clusters 
(e.g., [$\alpha$/Fe] = $+$0.3), then our overall photometric metallicities would become 
more metal-poor by $\sim$0.1 -- 0.2 dex.  This comparison is shown as a blue 
curve in Figure~\ref{fig:metcomp}.

Our two independent metallicity measurements generally agree with one another, 
and we have shown that the scatter in the spectroscopic metallicities results 
largely from poor $S/N$.  Therefore, we will adopt the photometric metallicities 
for the subsequent analysis.  We stress that the reported abundances are 
derived from a secure sample of member stars in each satellite.

\subsection{Abundance Distributions}\label{abundancedistributions}

The photometric metallicity distribution functions (MDFs) for the 
confirmed RGB stars in each of And~I, II, and III in displayed in 
Figure~\ref{fig:abunddist}.  The top panels show 
discrete histograms and the bottom panels show the cumulative distributions.  
The dashed line marks a fixed guide at [Fe/H] = $-$1.75 in both panels.  As 
was apparent from the CMDs, we confirm that all three of these dSphs are 
in fact metal-poor galaxies, with And~I being slightly more metal-rich 
relative to And~II and III.  We find that the internal abundance spread of 
all three galaxies is quite similar, differing by less than 0.1~dex.  Formally, 
we measure the mean metallicity, error in the mean, and dispersion to be 
[Fe/H]$_{\rm phot}$ = $-$1.45 $\pm$ 0.04 ($\sigma_{\rm [Fe/H]}$ = 0.37) for And~I, 
[Fe/H]$_{\rm phot}$ = $-$1.64 $\pm$ 0.04 ($\sigma_{\rm [Fe/H]}$ = 0.34) for And~II, and 
[Fe/H]$_{\rm phot}$ = $-$1.78 $\pm$ 0.04 ($\sigma_{\rm [Fe/H]}$ = 0.27) for And~III.  

Abundance analysis of And~VII, X, and XIV were presented in \cite{grebel99}, 
\cite{kalirai09}, and \cite{majewski07}.  Summarizing, the high resolution 
spectroscopic data for And~VII in \cite{guh00} did not yield an abundance measurement 
due to the low $S/N$ of the 18 confirmed giants at the Ca\,{\sc ii} triplet.  Therefore, 
we adopt And~VII's metallicity from the Keck/LRIS photometric study of \cite{grebel99}, 
who find [Fe/H]$_{\rm phot}$ = $-$1.4 $\pm$ 0.3 (the dispersion in metallicity was not 
measured).  For And~X, \cite{kalirai09} measured both a 
spectroscopic and photometric metallicity and found these to be in nice 
agreement.  They find that And~X is a metal-poor galaxy with [Fe/H]$_{\rm phot}$ = 
$-$1.93 $\pm$ 0.11, and exhibits a large metallicity dispersion of 
$\sigma_{\rm [Fe/H]}$ = 0.48.  For a distance modulus of 24.7, the the 
photometric metallicity of And~XIV is very low, [Fe/H] = $-$2.26 $\pm$ 0.05 (there 
is no metallicity dispersion measurement).  These results are also summarized in Table~2.

For And~I, II, and III, we can compare our metallicity results to the Da~Costa et~al.\ 
(1996; 2000; 2002) {\it HST}/WFPC2 study.  Da~Costa et~al.\ transformed their photometry from the 
native HST filters to Johnson-Cousins, and then compared the color of the RGB 
stars to a set of giant branches for Galactic globular clusters (i.e., 
alpha-enhanced populations).  As we noted earlier, their study of And~I is likely affected 
by the metal-rich stream debris that we eliminated from this sightline in 
\S\,\ref{membership}.  Interestingly, our mean metallicity is still identical to what 
they found, [Fe/H]$_{\rm phot}$ = $-$1.45 $\pm$ 0.20, suggesting that their value was in fact 
underestimated.  As expected, our dispersion is much smaller than their study, 
$\sigma_{\rm [Fe/H]}$ = 0.60.  Our sample only includes stars with [Fe/H] $<$ $-$0.92 given the 
detected substructure in this field, and, therefore, we would have eliminated 
the presence of any metal-rich tail in this dSph.  For And~II, we find a very 
similar abundance spread in the galaxy compared to the results of Da~Costa et~al.\ 
($\sigma_{\rm [Fe/H]}$ = 0.36), however our mean metallicity is more metal-poor than 
their study by $\sim$0.15 dex.  For And~III, our mean metallicity is consistent 
with the Da~Costa et~al.\ study within their larger error bar ([Fe/H]$_{\rm phot}$ 
= $-1.88\pm0.11$), however our internal metallicity dispersion is more than 2$\times$ 
larger than their measurement ($\sigma$ = 0.12).  

Recently, \cite{mcconnachie07} imaged And~II using the 
wide-field Subaru Suprime-Cam instrument to a photometric depth below the horizontal 
branch.  Their analysis suggests that the dSph contains two distinct 
components.  For the dominant extended component, they find an old population with 
[Fe/H] = $-$1.5 and $\sigma_{\rm [Fe/H]}$ = 0.28.  For the more concentrated inner 
region of the dSph, they find an intermediate aged (7 -- 10~Gyr) and metal-rich 
([Fe/H] = $-$1.2 and $\sigma_{\rm [Fe/H]}$ = 0.40) population.  Confirming this 
picture with our present spectroscopic data is difficult given the limited spatial 
coverage (e.g., lack of many stars near the center of the galaxy), however, we have 
now obtained five times as many spectroscopic measurements over the face of this 
galaxy.  These data, consisting of $\sim$500 individual radial velocity 
measurements of And~II members, will present a much cleaner view of the nature of 
And~II's stellar populations.

Overall, our results suggest that the chemical abundances and abundance spread (where 
available) of the brighter four M31 satellites in our sample are quite similar, and that 
there is no evidence suggesting the evolutionary histories of these galaxies were significantly 
different.  The two faintest satellites in this clean sample, And~X and XIV, are found to be 
more metal-poor than the brighter satellites.  Of course, a full analysis will require deeper 
photometric observations of the main-sequence turnoff morphology in each system, from which a 
full star formation history can be derived in conjunction with these metallicity 
measurements.  We also note that the metallicities of the M31 halo dSphs 
is similar to the mean metallicity of M31's stellar halo, determined by 
\cite{kalirai06a} and \cite{chapman06} to be [Fe/H] $\sim$ $-$1.5.


\section{Global Properties: \\ Comparing Milky Way and M31 dSphs} \label{globalproperties}

\subsection{An Inventory of the Milky Way and M31 dSphs} \label{inventory}

As we introduced earlier, wide field imaging surveys have recently uncovered many new dSph galaxies 
orbiting the Milky Way.  As rapidly as these systems are being discovered, different groups 
have targeted individual giant and dwarf stars with multiobject spectrographs to characterize 
the radial velocities, velocity dispersions, and spectroscopic abundances of these systems.  
For comparison to our M31 sample, we begin by considering only those Milky Way dSphs in which 
the intrinsic velocity dispersion has been resolved.  

Our baseline catalogue of properties for the classical Milky Way dSphs is drawn from \cite{irwin95} and 
the review by \cite{mateo98}, updated to reflect recent photometric and spectroscopic analysis 
on a galaxy by galaxy basis where available.  Our sample includes the Milky Way dSphs Carina, Draco, 
Fornax, Leo~I, Leo~II, Sculptor, Sextans, and Ursa~Minor, as well as the more distant Local Group 
dSphs Tucana and Cetus \citep{mcconnachie06,lewis07}.  For the core radii and central velocity 
dispersions, we adopt values from \cite{wolf09} who present a summary of the most updated 
results in their Table~1 (see full references in their study, including Walker, Mateo, \& Olszewski 
2009).  The [Fe/H] of most of these galaxies is updated from the 
\cite{mateo98} results based on new multiobject studies of larger numbers of stars.  For example, 
for Carina, Fornax, Sculptor, and Sextans we use the latest DART team results \citep{helmi06}, 
for Leo~I and Leo~II we use the \cite{koch07a} and \cite{koch07b} studies, and for Draco and 
Ursa~Minor we adopt the values in \cite{winnick03}.  If it is not directly measured in these studies, 
we have assumed an uncertainty of 10\% in the half light radius of each galaxy.  For the newly 
discovered lower luminosity Milky Way dSphs, our sample includes Bootes~I, Bootes~II, 
Canes Venatici~I, Canes Venatici~II, Coma Berenices, Hercules, Leo~IV, Leo~T, Segue~I, Ursa Major~I, 
and Ursa Major~II.  The baseline properties, such as luminosity and half-light radius, come from 
the summary table in \cite{wolf09} (see references therein and also the structural analysis in 
Martin, de~Jong, \& Rix~2008), velocity dispersions come from the Keck/DEIMOS spectroscopic 
study by \cite{simon07}, and abundances come from \cite{kirby08}.  The core radii of the dSphs 
is again taken from the King model fits in \cite{strigari08b}, with a 10\% assumed error.  We 
note that the core radius of Bootes~I is not known and so we have made an approximation such 
that $r_c$ $\sim$ $R_{e}$.  This can lead to a $\sim$25\% uncertainty \citep{martin07}, 
which has been factored into the calculation of dynamical masses (see below).  


\begin{figure*}
\begin{center}
\leavevmode
\includegraphics[width=13cm,angle=270, bb= 22 16 578 784]{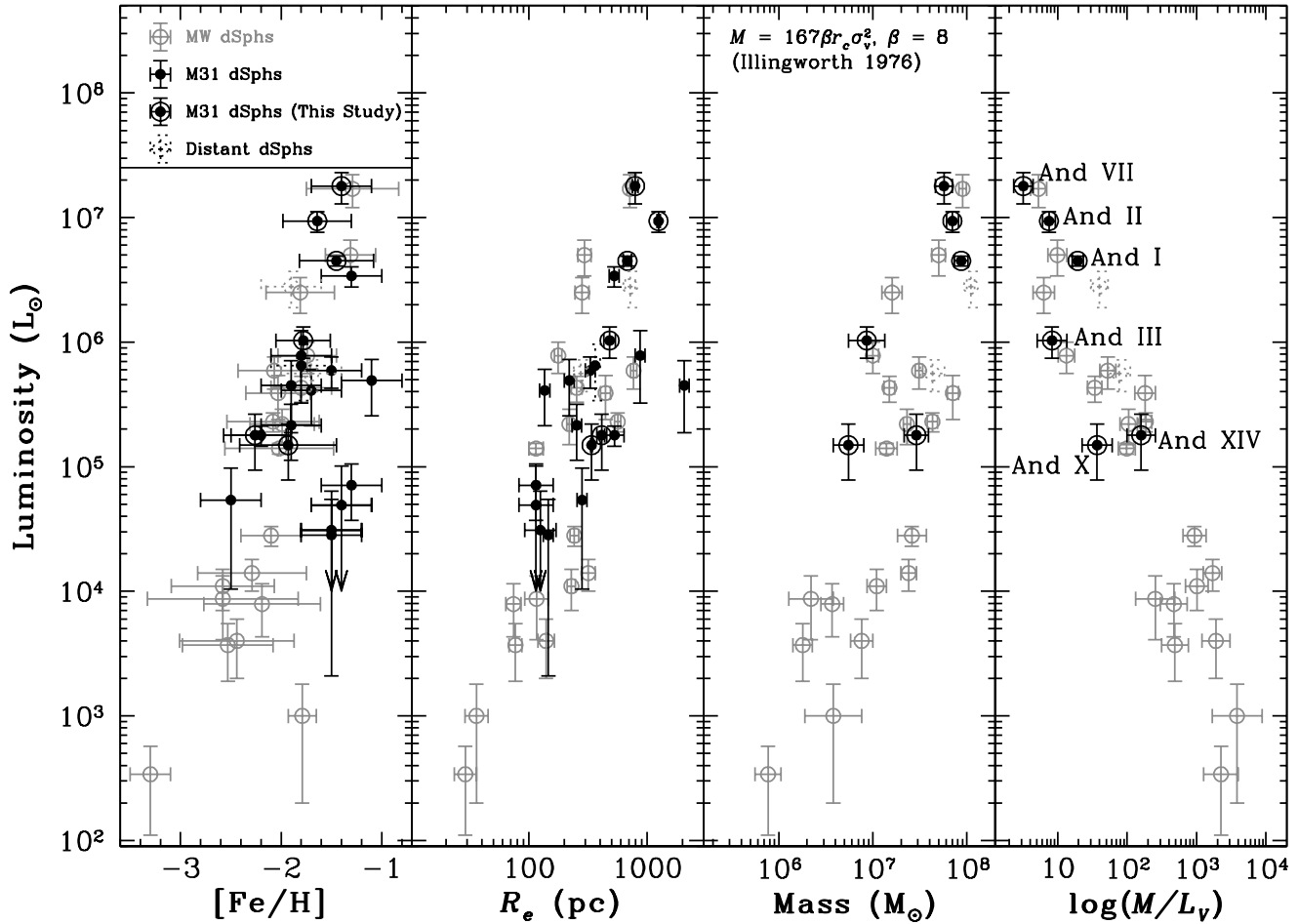}
\end{center}
\caption{The properties of M31 dSphs are compared to Milky Way dSphs of the same luminosity 
(see \S\,\ref{globalproperties} for references).  The Milky Way dSphs are illustrated with 
as grey points whereas the M31 points are darker.  The six M31 galaxies for 
which we have resolved intrinsic velocity dispersion measurements from the present study are 
dark encircled points.  Finally, the three distant Local Group dSphs, Cetus, Tucana, and 
And~XVIII are shown with dotted points and errorbars.  All results are summarized in Table~3.
The left panel illustrates the luminosity-metallicity relation, where we find excellent overlap 
between the M31 and Milky Way satellites, suggesting these systems have had similar chemical 
evolution histories.  Four of the faintest M31 dSphs are found to be signicantly more metal-rich 
than expected from the trend, and therefore may be systems that have experienced tidal stripping 
(i.e., they are underluminous for their metallicity).  The error bars in this plot represent the 
dispersion in metallicity in each galaxy.  For several dSphs lacking such measurements (mostly in 
M31), the dispersion has been arbitrarily set to 0.30~dex (see Table~3).  The middle-left panel 
illustrates the luminosity-size relation, where the size is represented by the 2D elliptical half-light 
radius.  Although the brightest M31 satellites are larger than Milky Way dSphs 
of the same luminosity (see also McConnachie \& Irwin 2006), we find that lower luminosity systems 
have similar sizes and therefore dynamical effects are unlikely to have truncated the sizes of 
these systems in a systematic way.  The masses and mass-to-light ratios of the M31 dSphs, as 
calculated using the simplified technique described in \S\,\ref{meanvelocity}, are illustrated in 
the two right panels.  The overall trend of decreasing total mass-to-light ratio with increasing 
luminosity is similar for the dSphs of each galaxy, although two of the faintest M31 dSphs in our 
sample appear to be less massive than their Milky Way counterparts.  Combined with 
their similar (or larger) sizes, there is mild evidence that the densities of these galaxies are 
lower.  This is verified by considering the more carefully calculated mass at the half light 
radius based on the new \cite{wolf09} method, as described in \S\,\ref{DMmasses} and presented in 
Figure~\ref{fig:masses}.}
\label{fig:properties}
\end{figure*}


For the 19 M31 dSphs, plus And XVIII which appears to be a background object, we adopt parameters primarily 
from the discovery papers summarized earlier as well as the analysis of \cite{mcconnachie06} for the brightest 
satellites.  The half light radii in \cite{mcconnachie06} are computed as geometric means, and so we have 
adjusted these to an elliptical half-light radius ($R_{e}$) by factoring in the measured ellipticity.  
The metallicity of And~V is taken from \cite{armandroff98} and for And~VI and VII is taken 
from \cite{grebel99}.  For the six satellites we've discussed in this paper, we adopt our new results based 
on the spectroscopically clean samples of stars.  Again, a 10\% error in the core radius is assumed (i.e., 
for the calculation of dynamical masses) if it has not been measured in the earlier papers.  As discussed 
in \S\,\ref{intro}, the limited kinematic data presented for And~IX and And~XII \citep{chapman05,chapman07,collins09}, 
And~XI and XIII \citep{collins09}, and And~XV and XVI \citep{letarte09} have large uncertainties and are inconclusive 
in yielding accurate dynamical masses, and so we ignore them in the present analysis.

We recalculate the dynamical mass of each of the Milky Way dSphs summarized above using the same (simplified) 
technique described in \S\,\ref{meanvelocity}, with updated velocity dispersion and core radius measurements 
from these studies.  More accurate masses of these Milky Way satellites at the half light radii are calculated 
in \cite{wolf09}.  A summary of all of these current properties, for both the Milky Way dSphs with internal 
velocity dispersion measurements, and for all of the M31 dSphs, is given in Table~3.


\subsection{Chemical Abundance Trends}\label{chemabundtrends}

Table~3 illustrates that the Milky Way and M31 each host $\sim$20 dSphs, and the bulk of 
these satellites have measured abundances and half light radii.  The intrinsic luminosity range 
of the M31 dSphs is of course smaller than the Milky Way satellites, however the two distributions 
overlap over a factor of 1000 in luminosity.  There exists a known luminosity-metallicity relation 
for Local Group dSphs extending down to the faintest systems (e.g., Mateo 1998; Grebel et~al.\ 2003; 
Kirby et~al.\ 2008), and the M31 satellites are found to track the brighter part of this relation 
remarkably well, e.g., from $L$ $\sim$ 10$^{5}$ to 10$^{7}$ $L_\odot$.  We illustrate this in the left 
panel of Figure~\ref{fig:properties}, where the darker encircled symbols represent the spectroscopically 
cleaned measurements from our present study.  Such a strong correlation demonstrates that the gross chemical 
evolution histories of these satellites are similar, and, if at all, similarly affected by dynamical 
interactions from their different host environments (see \S\,\ref{dynamicalinteractions}).  As summarized 
in Table~3, the intrinsic 
metallicity spreads of the brightest satellites in the Milky Way and M31 are also similar to one 
another, and as large as 0.3 -- 0.5 dex (the metallicity dispersion is indicated with horizontal 
error bars in Figure~\ref{fig:properties}).\footnote{A horizontal error bar of 0.30~dex has been 
plotted for any dSphs that lack a metallicity dispersion measurement -- See Table~3.}  These 
similarities indicate that the stars in these 
galaxies may have resulted from an extended (early) star formation history, with the earlier 
generation of stars polluting the interstellar medium for subsequent star formation bursts.  Yet, 
the trends of these overall processes must be very similar in the dSphs of the two hosts, and therefore 
minimally affected by external influences.  As we note below, the gravitational potential of the dSphs 
in the two hosts also appears to be similar.  Neither of these two findings, that there is a 
correlation between luminosity and metallicity or that the intrinsic abundance spreads are large, 
are true for globular clusters \citep{harris96}.

Interestingly, four of the least luminous M31 dSphs (And~XI, XII, XIII, and XX) are found to be 
significantly more metal-rich than expected given their luminosities.  These four systems, with 
$-$7.3 $<$ $M_{\rm V}$ $<$ $-$6.3, are found to be 0.5 -- 1~dex more metal-rich than a reasonable 
extrapolation of both the M31 population and from the Milky Way dSphs with similar luminosity.  
One possible interpretation of this trend is that these satellites were initially 
much more massive than present, and have subsequently experienced tidal stripping.  
Such a scenario could possibly preserve the mean metallicity of the galaxies, while at 
the same time, reduce their luminosities significantly.  The presence of a positive radial 
abundance gradient (which is not known) towards the center of any of these dSphs would 
enhance this effect, as the more metal-poor stars in the outskirts would be 
preferentially stripped away (e.g., Chou et~al.\ 2007).  Unfortunately, testing the tidal 
stripping possibility requires detailed kinematic data for many stars in these satellites, 
especially at larger radii.  Such a data set does not yet exist in these low-luminosity 
systems.  Similar to the M31 sample, the faint Milky Way satellite Bootes~II 
also deviates from the general trend, possibly suggesting a similar evolutionary history as 
outlined for these M31 satellites.  

\subsection{Physical Sizes}\label{sizes}

\cite{mcconnachie06} found the very interesting result that the sizes of M31's dSphs appear 
to be a factor of two larger than Milky Way dSphs of similar luminosity.  
The sample in their photometric study included some of M31's brightest satellites, 
And~I, II, III, V, VI, and VII (e.g., down to $M_{\rm V}$ = $-$9.6).  In Figure~\ref{fig:properties} 
({\it middle-left panel}), we extend this comparison by a factor of 20 down to a luminosity 
of $M_{\rm V}$ = $-$6.3 by including all 19 M31 dSphs from the discovery papers summarized 
earlier.  With the exception of the two newly found galaxies And XIX ($R_{e}$ = 2.07~kpc -- corrected from 
a geometric mean to an elliptical radius; McConnachie et~al.\ 2008), and And XXI ($R_{e}$ = 875~pc; Martin et~al.\ 
2009), the bulk of the newly found dSphs 
in M31 have sizes comparable to Milky Way counterparts of similar luminosity.  We also note that the 
two outliers in the new data are all among the furthest galaxies along the line of sight, and have 
similar (or lower) luminosities than counterparts with smaller sizes.  Therefore, the measurement of 
their sizes is more difficult given the fewer number of detected member stars.


\begin{figure}
\begin{center}
\leavevmode
\includegraphics[width=8.5cm]{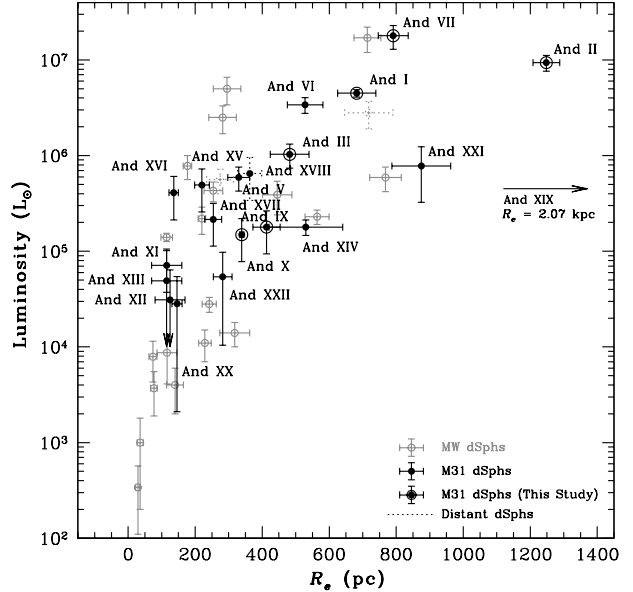}
\end{center}
\caption{The 2D elliptical half-light radii of Milky Way dSphs is compared to the M31 dSphs on a linear 
scale.  In cases where the uncertainties in the half light radii were not directly measured, 
we have assumed it to be 10\% in this plot.  The general trend of dSphs in the two large 
hosts indicates that the most luminous M31 satellites extend to larger sizes than the Milky 
Way dSphs (see also McConnachie \& Irwin 2006). However, for the bulk of the lower luminosity 
systems (e.g., $L$ $\lesssim$ 10$^6$~$L_\odot$), the two distributions significantly overlap.} 
\label{fig:sizes}
\end{figure}


The overall size distribution of the two satellite systems can be better seen on an expanded 
linear scale, as presented in Figure~\ref{fig:sizes}.  The distributions are found to 
significantly overlap between $L$ = 10$^{4}$ and 10$^{6}$~$L_\odot$.  In fact, at lower 
luminosities, it appears And~XI, XII, XIII, and XX all have systematically {\it smaller} half 
light radii than similar luminosity Milky Way galaxies such as Bootes~I, Hercules, Leo~T, and Ursa 
Major~I.  We note that the uncertainty in $R_{e}$ in both of these figures is arbitrarily taken 
to be 10\% in cases where the error has not been directly measured.  From this overall comparison, we can say 
that less than one-third of the M31 satellites are consistently larger than Milky Way counterparts 
of the same luminosity, and that this result is mostly limited to the brighter galaxies.  This evidence 
therefore does not strongly support the claim that the Milky Way has a stronger tidal field that 
truncates its dSphs at a smaller radius relative to M31, unless such a dynamical effect could be 
limited to the brightest satellites.

\subsection{Dark Matter Masses}\label{DMmasses}

The radial velocities from the SPLASH Survey presented in this paper allow us to measure 
dynamical masses for roughly one-third of the entire known population of M31's dSphs.  In 
Table~3, we have summarized the masses and mass-to-light ratios of these galaxies, based 
on the analysis presented in \S\,\ref{meanvelocity} (i.e., using the Illingworth 1976 
method).  We have also recalculated the masses of the Milky Way dSphs with resolved central 
velocity dispersions using the same technique.  

As we noted earlier, these mass measurements follow from a simple scaling relation 
involving the measured sizes and velocity dispersions of these galaxies, and assume an isotropic 
sphere in which mass follows light.  Depending on the quality of the observations, several 
Milky Way dSphs allow modeling of the individual velocities to constrain the overall mass profile 
of the satellites (see e.g., Strigari et~al.\ 2008b and Walker et~al.\ 2009). \cite{wolf09} 
have presented a new prescription to calculate the masses of such systems out to the half light 
radius, and show that these results agree well with with full dynamical modeling approach.  We 
summarize the masses calculated from this technique, both for Milky Way and M31 satellites, 
in the last column of Table~3.  As \cite{wolf09} show, the \cite{illingworth76} approximation 
systematically underpredicts the mass of dSphs and so the new $M_{1/2}$ are 
often similar, if not larger, than those values.  This newer analysis is discussed in more 
detail below.

In the two right panels of Figure~\ref{fig:properties}, we illustrate the total measured masses 
and mass-to-light ratios of Milky Way dSphs and the six M31 dSphs that are the focus of this 
paper.  Overall, the trend of increasing total mass of Milky Way dSphs with higher luminosity 
persists in the M31 dSphs.  The slope of the relation between the two families of dSphs 
is also very similar.  At the luminous end, our sample includes the two brightest M31 dSphs 
known, And~II and VII, both of which have a similar luminosity to the Galactic dSph Fornax.  
Formally, our measurements indicate that the total masses of these three dSphs is 
6 -- 9 $\times$ 10$^{7}$~$M_\odot$.  At slightly lower luminosities, 
we find that the $M/L_{\rm V}$ of And~I ($M_{\rm V}$ = $-$11.8) is higher by a factor of two 
compared to the Milky Way dSph Leo~I ($M_{\rm V}$ = $-$11.5), yet is lower by a factor of two 
compared to the luminous distant Local Group dSph Cetus ($M_{\rm V}$ = $-$11.3).  For our three 
fainter satellites, we find that the $M/L_{\rm V}$ of And~XIV is in nice agreement with the 
comparable Milky Way dSphs Draco, Canes Venatici I, and Leo~T.  However, for And~III and And~X, 
we find that the measured $M/L_{\rm V}$ is smaller by several factors compared to most Milky Way 
dSphs of comparable brightness, such as Draco, Sextans, Ursa Minor, Canes Venatici I, and 
Leo~T, and therefore these galaxies contain less dark matter than their Milky Way 
counterparts (see also Kalirai et~al.\ 2009).

For the luminous dSphs, the comparisons above demonstrate that, despite having a 
larger size than their Milky Way counterparts of the same luminosity, the 
masses of dSphs around each of the two hosts are similar.  At lower luminosities, 
we see mild evidence that two of the six M31 satellites in our sample have a lower 
mass than their Milky Way counterparts.  As these latter two satellites have half 
light radii that are similar, if not slightly larger, than Milky Way dSphs, 
this hints at an overall trend of the six M31 dSphs being less dense than Milky Way 
dSphs. 


\begin{figure}
\begin{center}
\leavevmode
\includegraphics[width=8.5cm]{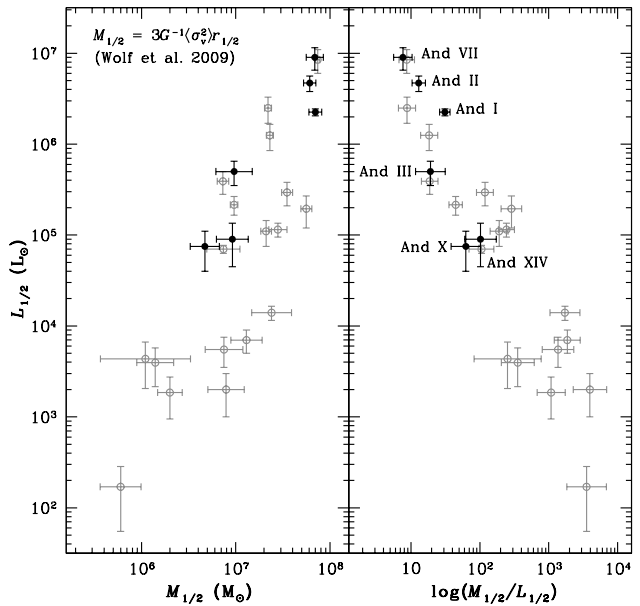}
\end{center}
\caption{The masses (left) and mass-to-light ratios (right) of M31 (black) and Milky Way (grey) 
dSphs at the half light radius, calculated as described in \cite{wolf09}.  The comparison of 
dSphs in the two hosts indicates that the lower luminosity M31 dSphs have dynamical masses and 
mass-to-light ratios that track the lower mass edge of the general relation for Milky Way 
satellites.  Although the dynamical masses of these satellites are generally smaller than 
their Milky Way counterparts, they have half light radii that are similar or larger (see 
Figures~\ref{fig:properties} and \ref{fig:sizes}).  These galaxies are therefore less 
dense than the Milky Way dSphs.}
\label{fig:masses}
\end{figure}


\cite{strigari08b} have recently constructed mass profiles for the Milky Way dSph 
population and surprisingly found that, over four orders of magnitude 
in luminosity, all of the satellites have approximately the same dark matter mass 
($\sim$10$^7$~$M_\odot$) within their central 300~pc.  The interpretation of a common 
mass scale for Milky Way dSphs is still unclear, as discussed by \cite{strigari08b}.  
For example, the common mass measurements may be indicative of a fundamental scale for 
dark matter clustering.  Correctly measuring whether such a scale extends to M31's 
dSphs, and if the absolute value is the same, will require both larger data sets 
(e.g., $\gtrsim$50 accurate radial velocities per dSph) and more sophisticated modeling 
techniques.  In the last column of Table~3, we have calculated the masses of each of the 
M31 dSphs and Milky Way satellites at the half light radius ($M_{1/2}$), using the new \cite{wolf09} 
technique (see \S\,\ref{accuratemasses}).  A comparison of $M_{1/2}$ between the Milky Way and 
M31 dSphs, at the same luminosity, verifies the findings above.  At the luminous end of the 
distribution, the $M_{1/2}$ are similar despite the M31 dSphs being larger, while at lower 
luminosities, the $M_{1/2}$ are smaller for the M31 dSphs (e.g., And~III, X, and XIV; see 
Figure~\ref{fig:masses}).  We will present the first results on the mass profiles of 
these M31 dSphs, and therefore the mass within a fixed physical radius, in Wolf et~al.\ 
(2010, in prep.).  

\subsection{The Bigger Picture: The Milky Way vs M31} \label{dynamicalinteractions}

Although the Milky Way and M31 are both generally characterized as large spiral galaxies, 
several differences between the two systems have been pointed out in 
recent studies such as the SPLASH Survey.  For example, surface brightness and abundance 
studies have shown that the metal-poor, stellar halo of M31 dominates the metal-rich 
inner spheroid at a radius of 20 -- 30~kpc, much farther out than the transition point in 
the Milky Way \citep{guh05,irwin05,kalirai06a,chapman06,ibata07,koch08}.  Additionally, the 
overall density of M31 at 20~kpc 
is much higher than the Milky Way \citep{reitzel98}, and M31 contains a higher fraction of 
intermediate age stars than the Milky Way at these radii \citep{brown07}.  Major differences 
among the population of satellites in the two systems have also long been known.  
The most luminous Milky Way satellites, the Small and Large Magellanic 
Clouds, are irregular type galaxies, of which M31 has none.  The most luminous M31 
satellites are the dwarf ellipticals NGC~205, NGC~187, and NGC~147, of which the Milky 
Way has none.  The overall population of globular clusters in M31 outnumbers the Milky 
Way by several times.

The differences between the Milky Way and M31 are likely tied to their past accretion 
histories.  Deep, wide field imaging surveys of M31 demonstrate a large amount of 
substructure both in the inner and outer halo \citep{ferguson02,ibata07,mcconnachie09}.  
Simulations of galaxy formation in the $\Lambda$CDM paradigm suggest that the inner halos 
of massive galaxies are built up quickly, and formed from just a few of the most massive, 
early accretion events (e.g., Diemand, Kuhlen, \& Madau 2007).  If the 
accretion history is responsible for the present day differences between the two large 
galaxies (e.g., M31 having experienced more interactions -- Hammer et~al.\ 2007), 
then it may be the case that the surviving satellites suffered enhanced stripping.  As 
we have discussed, the overall sizes of the M31 dSphs are systematically larger than 
their Milky Way counterparts at the luminous end (see also McConnachie \& Irwin 2006), 
and the dynamical masses of M31's dSphs are either similar or lower than comparable 
luminosity Milky Way dSphs. However, in order to ``puff'' these systems up into less 
dense objects, the interactions would have had to strip the dark matter halos within 
the visible radii of the galaxies, in which case the stars should also be stripped.  

An alternate mechanism for losing stars through stripping has also just been presented 
by \cite{donghia09}.  They simulate the interactions of dwarf disk galaxies with larger 
systems and find that the encounters can excite gravitational resonances.  They refer to 
this process as ``resonant stripping'' and show that the dwarf disk galaxies can transform 
to systems that look like present day dSphs (gas poor, dark matter dominated).  Resonant 
stripping allows for up to 80\% of the deeply embedded stellar component in satellite galaxies 
to be lost after a few Gyrs of interaction, while, at the same time, much less strongly affecting 
the surrounding dark matter halos.  Therefore the mass-to-light ratios of these galaxies, and 
the sizes of their stellar distribution, can be altered depending on the details of the 
interaction history (e.g., prograde vs retrograde orbits and the degree and timescale of the 
encounters).  Unfortunately, such an effect is difficult, if not impossible, to measure from the 
limited angular spread of our radial velocities, which are not well extended in large 
numbers to the tidal radius of each satellite.  A thorough investigation of the tidal interaction 
of these satellites with M31, and specifically relating such interactions to those of 
similar luminosity Milky Way dSphs, will also require an understanding of the orbits 
of these galaxies.  Tangential velocities are currently only available for a few of 
the most luminous Milky Way satellites, and for none of the M31 dSphs.  


\begin{figure}
\begin{center}
\leavevmode
\includegraphics[width=8.5cm]{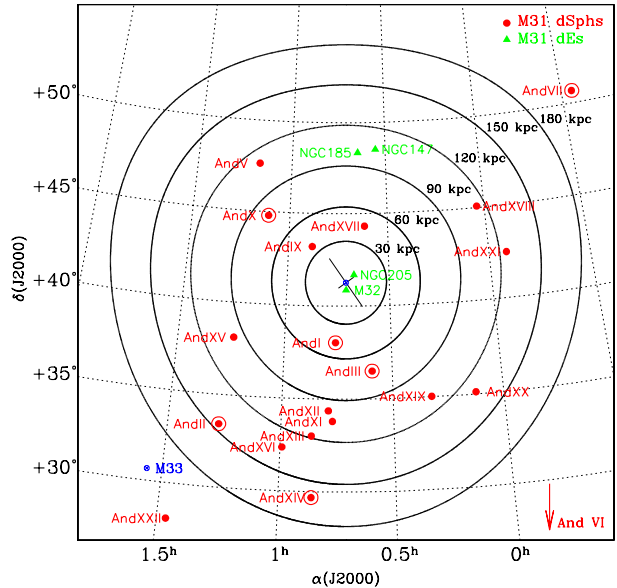}
\end{center}
\caption{The location of M31's 18 dSphs relative to the center of the galaxy 
is shown as filled red circles.  The six dSphs comprising this study are 
also indicated with larger open circles, and clearly span a wide range in 
distance from M31 and azimuth.  As described in the text, the bias of satellites 
in the south results from incomleteness in the imaging surveys.}\label{fig:locationdSphs}
\end{figure}


To first order, we can investigate the projected distribution of M31's 19 dSphs 
relative to the center of the galaxy, as shown in Figure~\ref{fig:locationdSphs}.  
The dSph distribution spans over 30 degrees in right ascension and declination on 
this Figure (red circles).  The concentric annuli are each 
separated by 30~kpc, and extend outward to 180~kpc from M31.  The small lines at 
the center of the figure mark the size of M31's optical disk.  Also shown are the 
locations of M33, and M31's 4 dwarf ellipticals (green triangles).  First, this figure 
demonstrates that there is a clear absence of M31 satellites in the 
northern regions of the galaxy.  The cause of this is an observational bias related 
to the location of deep imaging surveys that have thus far only explored more southern 
latitudes (e.g., Ibata et~al.\ 2007 and McConnachie et~al.\ 2009).  These surveys, 
e.g., PAandAS, are only now beginning to target other areas of the 
galaxy, and will surely uncover additional dSphs in M31, as will Pan-STARRS (Kaiser 
et~al.\ 2002).  The four recent discoveries of And~XVIII, And~XIX, And~XX, and 
And~XXI in the western hemisphere of M31 reflect the extention of the initial CFHT 
MegaCam survey \citep{mcconnachie08,martin09}.

The six satellites in our study are shown with enclosed open circles in 
Figure~\ref{fig:locationdSphs}.  These dSphs span an appreciable 
range in radius and azimuth relative to M31.  And I, III, and X sample 
intermediate radii in the halo, from $R$ = 45 -- 75~kpc, whereas And~II,
VII, and XIV are all located in the outskirts of M31's stellar halo, 
$R \gtrsim$ 150~kpc.  In this context, one or more of the three inner 
satellites may represent close-in bodies that were captured long ago, whereas 
the outer group could be, in part, dSphs that are just now starting to fall into M31's 
potential.  Evidence of such falling in galaxies is also supported by 
the radial velocity of And~XIV as discussed earlier (see also Majewski 
et~al.\ 2007), and also in the case of And~XII \citep{chapman07}.  However, 
both sets of galaxies, near and far from M31's center, include satellites that 
are systematically larger than Milky Way dSphs, as we showed in 
Figure~\ref{fig:properties}.  

Interestingly, the three lower luminosity M31 satellites And~XI, XII, and XIII 
appear to be smaller than a Milky Way counterpart of similar luminosity (Bootes~I), 
{\it and} are also located in a similar projected position in M31.  Kinematics of these 
galaxies will reveal if they are in fact moving at similar radial velocities, 
possibly suggesting an infalling group (e.g., see Li \& Helmi 2008; D'Onghia~2008).  
The dark matter masses of these satellites, in relation to the general trend of 
Milky Way vs M31 dSphs, will help gauge the degree to which external dynamical 
influences may have played a role in shaping the internal kinematics.  More 
generally, the future of resolved spectroscopy of M31 dSphs is sure to shed 
light on both the intrinsic properties of the satellites and on the galaxy 
formation and evolution processes that shape these properties.

\section{Conclusions} \label{conclusion}

We present the first systematic large scale spectroscopic survey of M31's dSphs 
as a part of the SPLASH Survey, a detailed study of And~I, II, III, VII, X, and 
XIV.  The fundamental properties of these galaxies are established from a sample 
of confirmed member stars, including the mean radial velocity, velocity dispersion, 
mean abundance, abundance spread, and total (stellar + dark matter) mass.  The kinematics indicate that 
the radial velocity and intrinsic velocity dispersion of the six satellites are 
$v_{\rm rad}=-375.8\pm1.4$~km~s$^{-1}$ and $\sigma_{v}=10.6\pm1.1$~km~s$^{-1}$ for And~I, 
$v_{\rm rad}=-193.6\pm1.0$~km~s$^{-1}$ and $\sigma_{v}=7.3\pm0.8$~km~s$^{-1}$ for And~II, 
$v_{\rm rad}=-345.6\pm1.8$~km~s$^{-1}$ and $\sigma_{v}=4.7\pm1.8$~km~s$^{-1}$ for And~III, 
$v_{\rm rad}=-309.4\pm2.3$~km~s$^{-1}$ and $\sigma_{v}=9.7\pm1.6$~km~s$^{-1}$ for And~VII, 
$v_{\rm rad}=-168.3\pm1.2$~km~s$^{-1}$ and $\sigma_{v}=3.9\pm1.2$~km~s$^{-1}$ for And~X, and 
$v_{\rm rad}=-481.0\pm1.2$~km~s$^{-1}$ and $\sigma_{v}=5.4\pm1.3$~km~s$^{-1}$ for And~XIV. 

Both photometric and spectroscopic metallicities are measured for all confirmed member 
stars in each galaxy, and the two independent measurements agree nicely with one 
another.  We find that the six dSphs are metal-poor systems with a  mean iron abundance 
of [Fe/H]$_{\rm phot}$ = $-1.45\pm0.04$ ($\sigma_{\rm [Fe/H]}$ = 0.37) for And~I,
[Fe/H]$_{\rm phot}$ = $-1.64\pm0.04$ ($\sigma_{\rm [Fe/H]}$ = 0.34) for And~II,
[Fe/H]$_{\rm phot}$ = $-1.78\pm0.04$ ($\sigma_{\rm [Fe/H]}$ = 0.27) for And~III, 
[Fe/H]$_{\rm phot}$ = $-1.40\pm0.30$ for And~VII, 
[Fe/H]$_{\rm phot}$ = $-1.93\pm0.11$ ($\sigma_{\rm [Fe/H]}$ = 0.248) for And~X, and 
[Fe/H]$_{\rm phot}$ = $-2.26\pm0.05$ for And~XIV (where two of the satellites lack a metallicity 
dispersion measurement).  

The dynamical mass of each satellite is estimated to first order using the \cite{illingworth76} 
formalism.  More accurate mass measurements at the half-light radius of each galaxy are also 
calculated using the methods described in \cite{wolf09}.  As summarized in Table~2 and 3, these 
yield half-light radius mass-to-light ratios of $M_{1/2}$/$L_{1/2}$ = 31 $\pm$ 6 for And~I, 
13 $\pm$ 3 for And~II, 19 $\pm$ 12 for And~III, 7.7 $\pm$ 2.8 for And~VII, 63 $\pm$ 40 for And~X, 
and 102 $\pm$ 71 for And~XIV.  These calculations are found to agree nicely with masses measured 
through full modeling of the individual radial velocities assuming the dark matter follows a five 
parameter density profile.

The comparison of the overall properties of M31's dSphs with Milky Way dSphs of similar luminosity 
indicates some interesting similarities and differences.  For example, we find that the luminosity -- 
metallicity relation is very similar between $L$ $\sim$ 10$^{5}$ -- 10$^{7}$ $L_\odot$, suggesting 
that the chemical evolution histories of each group of dSphs is similar.  The lowest luminosity 
M31 dSphs appear to deviate from the relation, possibly suggesting tidal stripping.  The previous 
finding of McConnachie \& Irwin (2006) indicating that the brightest M31 dSphs are systematically 
larger than Milky Way counterparts of the same luminosity does not persist cleanly to lower 
luminosities.  We find that the sizes of dSphs in the two hosts significantly overlap for $L$ 
= 10$^{4}$ -- 10$^{6}$~$L_\odot$, and that four of the faintest M31 dSphs are {\it smaller} than 
Milky Way satellites of similar luminosity.  The first dynamical mass measurement comparisons 
between the M31 and Milky Way dSphs hints that the M31 satellites are systematically less dense 
than Milky Way counterparts.


\acknowledgments

We thank Eva Grebel, Steve Vogt, and Dan Zucker for their valuable contribution 
to the And~VII and X observations.  JSK's research is supported in part by a grant from the STScI Director's 
Discretionary Research Fund, and was supported by NASA through Hubble Fellowship 
grant HF-01185.01-A, awarded by the Space Telescope Science Institute, which is 
operated by the Association of Universities for Research in Astronomy, Incorporated, 
under NASA contract NAS5-26555.  This project was also supported by NSF grants 
AST-0307966, AST-0507483, AST-0607852, and AST-0808133 and NASA/STScI grant 
GO-10265.02 (JSK, PG, KMG, and ENK), an NSF Graduate Fellowship (KMG), a Hubble 
Fellowship grant (HST-HF-01233.01) awarded by STScI (ENK), and NSF grants AST-0307842, 
AST-0307851, and AST-0607726, NASA/JPL contract 1228235, the David and Lucile Packard 
Foundation, and The F.~H.~Levinson Fund of the Peninsula Community Foundation 
(SRM, RJP, RLB).


\begin{sidewaystable*}\scriptsize
\begin{center}
\caption{}
\begin{tabular}{lcccccccccccc}
\hline
\hline
\multicolumn{1}{l}{Galaxy} & \multicolumn{1}{c}{$M_{\rm V}$} & \multicolumn{1}{c}{$L_{\rm V}$ ($L_\odot$)} &
\multicolumn{1}{c}{$d$ (kpc)} & \multicolumn{1}{c}{$\sigma_v$ (km~s$^{-1}$)} &
\multicolumn{1}{c}{$r_{c}$ (pc)\tablenotemark{1}} & \multicolumn{1}{c}{$R_{e}$ (pc)\tablenotemark{1}} & \multicolumn{1}{c}{$r_{1/2}$ (pc)\tablenotemark{1}} & 
\multicolumn{1}{c}{[Fe/H]}  & \multicolumn{1}{c}{$\sigma_{[Fe/H]}$} & \multicolumn{1}{c}{$M$ ($M_\odot$)\tablenotemark{2}} & 
\multicolumn{1}{c}{$M$/$L$ ($M_\odot$/$L_\odot$)\tablenotemark{2}} & \multicolumn{1}{c}{$M_{1/2}$ ($M_\odot$)\tablenotemark{3}} \\
\hline
{\bf M31 dSphs -- This Study} \\                                                                                                                                                                                                                 
{\bf And I	  } & {\bf  $-$11.8} & {\bf   (4.5 $\pm$ 0.4) $\times$ 10$^{6}$} & {\bf    745 $\pm$  24} & {\bf  10.6 $\pm$  1.1} & {\bf   580 $\pm$  60} & {\bf   682 $\pm$ 57 } & {\bf   900 $\pm$  75}  & {\bf  $-$1.45 $\pm$ 0.04} & {\bf  0.37} & {\bf  (8.7 $\pm$ 1.6) $\times$ 10$^{7}$} & {\bf    19 $\pm$ 4  } & {\bf  (7.0 $\pm$ 1.2) $\times$ 10$^{7}$} \\ 
{\bf And II       } & {\bf  $-$12.6} & {\bf   (9.4 $\pm$ 1.8) $\times$ 10$^{6}$} & {\bf    652 $\pm$  18} & {\bf   7.3 $\pm$  0.8} & {\bf   990 $\pm$  40} & {\bf  1248 $\pm$ 40 } & {\bf   1659 $\pm$  53} & {\bf  $-$1.64 $\pm$ 0.04} & {\bf  0.34} & {\bf  (7.0 $\pm$ 1.1) $\times$ 10$^{7}$} & {\bf   7.5 $\pm$ 1.8} & {\bf  (6.1 $\pm$ 1.0) $\times$ 10$^{7}$} \\
{\bf And III	  } & {\bf  $-$10.2} & {\bf   (1.0 $\pm$ 0.3) $\times$ 10$^{6}$} & {\bf    749 $\pm$  24} & {\bf   4.7 $\pm$  1.8} & {\bf   290 $\pm$  40} & {\bf   482 $\pm$ 58 } & {\bf   638 $\pm$  77}  & {\bf  $-$1.78 $\pm$ 0.04} & {\bf  0.27} & {\bf  (8.6 $\pm$ 4.8) $\times$ 10$^{6}$} & {\bf   8.3 $\pm$ 5.2} & {\bf  (9.6 $\pm$ 5.4) $\times$ 10$^{6}$} \\
{\bf And VII	  } & {\bf  $-$13.3} & {\bf   (1.8 $\pm$ 0.5) $\times$ 10$^{7}$} & {\bf    763 $\pm$  35} & {\bf   9.7 $\pm$  1.6} & {\bf   450 $\pm$  20} & {\bf   791 $\pm$ 45 } & {\bf   1050 $\pm$  60} & {\bf  $-$1.4  $\pm$ 0.30} & {\bf  \nodata} & {\bf  (5.7 $\pm$ 1.3) $\times$ 10$^{7}$} & {\bf   3.2 $\pm$ 1.2} & {\bf  (6.9 $\pm$ 1.6) $\times$ 10$^{7}$} \\
{\bf And X	  } & {\bf   $-$8.1} & {\bf   (1.5 $\pm$ 0.7) $\times$ 10$^{5}$} & {\bf    701 $\pm$  37} & {\bf   3.9 $\pm$  1.2} & {\bf   271 $\pm$  27} & {\bf   339 $\pm$ 6 } &  {\bf   448 $\pm$  8}   & {\bf  $-$1.93 $\pm$ 0.11} & {\bf  0.48} & {\bf  (5.5 $\pm$ 2.5) $\times$ 10$^{6}$} & {\bf   37  $\pm$ 24} & {\bf  (4.7 $\pm$ 2.0) $\times$ 10$^{6}$} \\
{\bf And XIV\tablenotemark{4}} & {\bf   $-$8.3} & {\bf   (1.8 $\pm$ 0.9) $\times$ 10$^{5}$} & {\bf    871 $\pm$  87} & {\bf   5.4 $\pm$  1.3} & {\bf 734 $\pm$  73} & {\bf 413 $\pm$ 41 } & {\bf 461 $\pm$ 155 } & {\bf  $-$2.26 $\pm$ 0.05} & {\bf  0.31} & {\bf  (2.9 $\pm$ 1.0) $\times$ 10$^{7}$} & {\bf   160 $\pm$ 95} & {\bf  (9.2 $\pm$ 4.4) $\times$ 10$^{6}$} \\
\hline
\hline
{\bf Other M31 dSphs} \\                                                                                                                                                                                                                 
And V	                   & $-$9.6  &  (5.9 $\pm$ 1.7) $\times$ 10$^{5}$ &   774 $\pm$  28 &  \nodata  &  280 $\pm$  40 &  330 $\pm$ 33  & \nodata & $-$1.5  \nodata    & \nodata & \nodata & \nodata & \nodata \\
And VI	                   & $-$11.5 &  (3.4 $\pm$ 0.6) $\times$ 10$^{6}$ &   783 $\pm$  25 &  \nodata  &  480 $\pm$  60 &  528 $\pm$ 53  & \nodata & $-$1.3  $\pm$ 0.30 & \nodata & \nodata & \nodata & \nodata \\
And IX	                   &  $-$8.3 &  (1.8 $\pm$ 0.3) $\times$ 10$^{5}$ &   787 $\pm$ 157 &  \nodata  &  309 $\pm$  31 &  530 $\pm$ 110 & \nodata & $-$2.2 \nodata     & \nodata & \nodata & \nodata & \nodata \\
And XI\tablenotemark{5}	   &  $-$7.3 &  (7.1 $\pm$ 3.4) $\times$ 10$^{4}$ &   783 $\pm$  78 &  \nodata  &  \nodata       &  115 $\pm$ 45  & \nodata & $-$1.3 \nodata     & \nodata & \nodata & \nodata & \nodata \\
And XII\tablenotemark{5}   &  $-$6.4 &  (3.1 $\pm$ 3.3) $\times$ 10$^{4}$ &   783 $\pm$  78 &  \nodata  &  \nodata       &  125 $\pm$ 45  & \nodata & $-$1.5 \nodata     & \nodata & \nodata & \nodata & \nodata \\
And XIII\tablenotemark{5}  &  $-$6.9 &  (4.9 $\pm$ 5.2) $\times$ 10$^{4}$ &   783 $\pm$  78 &  \nodata  &  \nodata       &  115 $\pm$ 45  & \nodata & $-$1.4 \nodata     & \nodata & \nodata & \nodata & \nodata \\
And XV\tablenotemark{6}	   &  $-$9.4 &  (4.9 $\pm$ 2.4) $\times$ 10$^{5}$ &   630 $\pm$  60 &  \nodata  &  167 $\pm$ 17  &  220 $\pm$ 22  & \nodata & $-$1.1 \nodata     & \nodata & \nodata & \nodata & \nodata \\
And XVI\tablenotemark{6}   &  $-$9.2 &  (4.1 $\pm$ 2.0) $\times$ 10$^{5}$ &   525 $\pm$  50 &  \nodata  &   98 $\pm$ 10  &  136 $\pm$ 14  & \nodata & $-$1.7 \nodata     & \nodata & \nodata & \nodata & \nodata \\
And XVII\tablenotemark{6}  &  $-$8.5 &  (2.2 $\pm$ 1.0) $\times$ 10$^{5}$ &   794 $\pm$  40 &  \nodata  &  \nodata       &  254 $\pm$ 25  & \nodata & $-$1.9 \nodata     & \nodata & \nodata & \nodata & \nodata \\
And XIX	                   &  $-$9.3 &  (4.5 $\pm$ 2.6) $\times$ 10$^{5}$ &   933 $\pm$  61 &  \nodata  &  \nodata       & 2065 $\pm$ 206 & \nodata & $-$1.9  $\pm$ 0.1  & \nodata & \nodata & \nodata & \nodata \\
And XX                     &  $-$6.3 &  (2.8 $\pm$ 2.6) $\times$ 10$^{4}$ &   802 $\pm$ 197 &  \nodata  &  \nodata       &  146 $\pm$ 15  & \nodata & $-$1.5  $\pm$ 0.1  & \nodata & \nodata & \nodata & \nodata \\
And XXI                    &  $-$9.9 &  (7.8 $\pm$ 4.6) $\times$ 10$^{5}$ &   859 $\pm$  51 &  \nodata  &  \nodata       &  875 $\pm$ 88  & \nodata & $-$1.8 \nodata     & \nodata & \nodata & \nodata & \nodata \\
And XXII\tablenotemark{4}  &  $-$7.0 &  (5.4 $\pm$ 4.4) $\times$ 10$^{4}$ &  1033 $\pm$ 103 &  \nodata  &  \nodata       &  282 $\pm$ 28  & \nodata & $-$2.5 \nodata     & \nodata & \nodata & \nodata & \nodata \\
\hline
{\bf Classical MW dSphs} \\
Carina	   &  $-$9.3 &  (4.3 $\pm$ 1.0) $\times$ 10$^{5}$ &  105 $\pm$  2  & 6.4  $\pm$ 0.2 &  269 $\pm$ 37  & 254 $\pm$ 28 & 334 $\pm$  37  & $-$1.80 $\pm$ 0.02 & 0.30 & (1.5 $\pm$ 0.2) $\times$ 10$^{7}$  &   34  $\pm$  9  & (9.6 $\pm$ 0.9) $\times$ 10$^{6}$ \\
Draco	   &  $-$8.6 &  (2.2 $\pm$ 0.7) $\times$ 10$^{5}$ &   76 $\pm$  5  & 10.1 $\pm$ 0.5 &  169 $\pm$ 11  & 220 $\pm$ 11 & 291 $\pm$  14  & $-$1.99 $\pm$ 0.04 & 0.32 & (2.3 $\pm$ 0.2) $\times$ 10$^{7}$  &   105 $\pm$ 35  & (2.1 $\pm$ 0.3) $\times$ 10$^{7}$ \\
Fornax	   &  $-$13.3 & (1.7 $\pm$ 0.5) $\times$ 10$^{7}$ &  147 $\pm$  3  & 10.7 $\pm$ 0.2 &  586 $\pm$ 53  & 714 $\pm$ 40 & 944 $\pm$  53  & $-$1.29 $\pm$ 0.02 & 0.46 & (9.0 $\pm$ 0.8) $\times$ 10$^{7}$  &   5.3 $\pm$ 1.6 & (7.4 $\pm$ 0.4) $\times$ 10$^{7}$ \\
Leo I	   &  $-$11.9 & (5.0 $\pm$ 1.6) $\times$ 10$^{6}$ &  254 $\pm$ 18  & 9.0  $\pm$ 0.4 &  459 $\pm$ 77  & 295 $\pm$ 49 & 388 $\pm$  64  & $-$1.31 $\pm$ 0.02 & 0.25 & (5.0 $\pm$ 0.9) $\times$ 10$^{7}$  &   9.9 $\pm$ 3.6 & (2.2 $\pm$ 0.2) $\times$ 10$^{7}$ \\
Leo II	   &  $-$9.9 &  (7.8 $\pm$ 2.2) $\times$ 10$^{5}$ &  233 $\pm$ 15  & 6.6  $\pm$ 0.5 &  179 $\pm$ 17  & 177 $\pm$ 13 & 233 $\pm$  17  & $-$1.74 $\pm$ 0.02 & 0.29 & (1.0 $\pm$ 0.2) $\times$ 10$^{7}$  &   13  $\pm$  4  & (7.3 $\pm$ 1.1) $\times$ 10$^{6}$ \\
Sculptor   &  $-$11.2 & (2.5 $\pm$ 0.8) $\times$ 10$^{6}$ &   86 $\pm$  5  & 9.0  $\pm$ 0.2 &  145 $\pm$ 41  & 282 $\pm$ 41 & 375 $\pm$  54  & $-$1.81 $\pm$ 0.02 & 0.34 & (1.6 $\pm$ 0.5) $\times$ 10$^{7}$  &   6.3 $\pm$ 2.7 & (2.3 $\pm$ 0.2) $\times$ 10$^{7}$ \\
Sextans	   &  $-$9.6 &  (5.9 $\pm$ 1.7) $\times$ 10$^{5}$ &   96 $\pm$  3  & 7.1  $\pm$ 0.3 &  461 $\pm$ 35  & 768 $\pm$ 47 & 1019 $\pm$ 62  & $-$2.07 $\pm$ 0.03 & 0.36 & (3.1 $\pm$ 0.3) $\times$ 10$^{7}$  &   53  $\pm$  16 & (3.5 $\pm$ 0.5) $\times$ 10$^{7}$ \\
Ursa Minor &  $-$9.2 &  (3.9 $\pm$ 1.5) $\times$ 10$^{5}$ &   77 $\pm$  4  & 11.5 $\pm$ 0.6 &  401 $\pm$ 51  & 445 $\pm$ 44 & 588 $\pm$  58  & $-$2.03 $\pm$ 0.04 & 0.32 & (7.1 $\pm$ 1.0) $\times$ 10$^{7}$  &  182  $\pm$  75 & (5.6 $\pm$ 0.8) $\times$ 10$^{7}$ \\
\hline
{\bf Lower Lum MW dSphs} \\
Bootes I\tablenotemark{7}   &  $-$6.3 &  (2.8 $\pm$ 0.5) $\times$ 10$^{4}$ &   66 $\pm$  3  &  9.0 $\pm$ 2.2 &  242 $\pm$  61 &  242 $\pm$ 21 & 322 $\pm$ 28 & $-$2.10 $\pm$ 0.30 & \nodata  & (2.6 $\pm$ 1.1) $\times$ 10$^{7}$ &   935 $\pm$  433 & (2.4 $\pm$ 1.5) $\times$ 10$^{7}$ \\
Bootes II                   &  $-$2.7 &  (1.0 $\pm$ 0.8) $\times$ 10$^{3}$ &   42 $\pm$  8  & 10.5 $\pm$ 7.4 &   26 $\pm$  3 &   36 $\pm$  9  & \nodata      & $-$1.79 $\pm$ 0.05 & 0.14    &  (3.8 $\pm$ 3.8) $\times$ 10$^{6}$ &  3830 $\pm$ 4909 & \nodata \\
Canes Venatici I	    &  $-$8.6 &  (2.3 $\pm$ 0.4) $\times$ 10$^{5}$ &  218 $\pm$ 10  &  7.6 $\pm$ 0.5 &  554 $\pm$  55 &  564 $\pm$ 36 & 750 $\pm$ 48 & $-$2.08 $\pm$ 0.02 & 0.46    & (4.3 $\pm$ 0.5) $\times$ 10$^{7}$  &   186 $\pm$   41 & (2.8 $\pm$ 0.7) $\times$ 10$^{7}$ \\
Canes Venatici II	    &  $-$4.9 &  (7.9 $\pm$ 3.7) $\times$ 10$^{3}$ &  160 $\pm$  5  &  4.6 $\pm$ 1.0 &  132 $\pm$  13 &   74 $\pm$ 12 & 97 $\pm$ 16  & $-$2.19 $\pm$ 0.05 & 0.58    & (3.7 $\pm$ 1.2) $\times$ 10$^{6}$  &   472 $\pm$  264 & (1.4 $\pm$ 0.8) $\times$ 10$^{6}$ \\
Coma Berenices              &  $-$4.1 &  (3.7 $\pm$ 1.8) $\times$ 10$^{3}$ &   44 $\pm$  4  &  4.6 $\pm$ 0.8 &   64 $\pm$  6 &   77 $\pm$ 10  & 100 $\pm$ 13 & $-$2.53 $\pm$ 0.05 & 0.45    & (1.8 $\pm$ 0.5) $\times$ 10$^{6}$  &   489 $\pm$  271 & (2.0 $\pm$ 0.7) $\times$ 10$^{6}$ \\
Hercules                    &  $-$5.3 &  (1.1 $\pm$ 0.4) $\times$ 10$^{4}$ &  133 $\pm$  6  &  5.1 $\pm$ 0.9 &  321 $\pm$  32 &  229 $\pm$ 19 & 305 $\pm$ 26 & $-$2.58 $\pm$ 0.04 & 0.51    & (1.1 $\pm$ 0.3) $\times$ 10$^{7}$  &  1014 $\pm$  459 & (7.5 $\pm$ 4.4) $\times$ 10$^{6}$ \\
Leo IV	                    &  $-$5.0 &  (8.7 $\pm$ 4.5) $\times$ 10$^{3}$ &  160 $\pm$ 15  &  3.3 $\pm$ 1.7 &  152 $\pm$  15 &  116 $\pm$ 30 & 151 $\pm$ 39 & $-$2.58 $\pm$ 0.08 & 0.75    & (2.2 $\pm$ 1.6) $\times$ 10$^{6}$  &   254 $\pm$  230 & (1.1 $\pm$ 2.2) $\times$ 10$^{6}$ \\
Leo T                       &  $-$8.1 &  (1.4 $\pm$ 0.1) $\times$ 10$^{5}$ &  407 $\pm$ 38  &  7.8 $\pm$ 1.6 &  170 $\pm$  17 &  115 $\pm$ 17 & 152 $\pm$ 21 & $-$2.02 $\pm$ 0.05 & 0.54    & (1.4 $\pm$ 0.4) $\times$ 10$^{7}$  &    99 $\pm$   32 & (7.4 $\pm$ 3.7) $\times$ 10$^{6}$ \\
Segue I                     &  $-$1.5 &  (3.4 $\pm$ 2.3) $\times$ 10$^{2}$ &   23 $\pm$  2  &  4.3 $\pm$ 1.1 &   31 $\pm$   3 &   29 $\pm$  7 & 38 $\pm$ 9   & $-$3.30 $\pm$ 0.20 & \nodata & (7.7 $\pm$ 2.9) $\times$ 10$^{5}$  &  2252 $\pm$ 1742 & (6.0 $\pm$ 3.9) $\times$ 10$^{5}$ \\
Ursa Major I                &  $-$5.6 &  (1.4 $\pm$ 0.4) $\times$ 10$^{4}$ &   97 $\pm$  4  &  7.6 $\pm$ 1.0 &  308 $\pm$  31 &  318 $\pm$ 45 & 416 $\pm$ 58 & $-$2.29 $\pm$ 0.04 & 0.54    & (2.4 $\pm$ 0.5) $\times$ 10$^{7}$  &  1698 $\pm$  603 & (1.3 $\pm$ 0.6) $\times$ 10$^{7}$ \\
Ursa Major II               &  $-$4.2 &  (4.0 $\pm$ 2.0) $\times$ 10$^{3}$ &   32 $\pm$  4  &  6.7 $\pm$ 1.4 &  127 $\pm$  13 &  140 $\pm$ 25 & 184 $\pm$ 33 & $-$2.44 $\pm$ 0.06 & 0.57    & (7.6 $\pm$ 2.4) $\times$ 10$^{6}$  &  1904 $\pm$ 1222 & (7.9 $\pm$ 4.4) $\times$ 10$^{6}$ \\
\hline                                                                                                                                                                                                                             
{\bf Distant LG dSphs} \\
Cetus                       &  $-$11.3 & (2.8 $\pm$ 0.9) $\times$ 10$^{6}$ &  755 $\pm$ 23 & 17.0 $\pm$ 0.2 &  290 $\pm$ 29 &  718 $\pm$ 72 & \nodata &  $-$1.9 \nodata & \nodata & (1.1 $\pm$ 0.1) $\times$ 10$^{8}$ &   40 $\pm$ 14 & \nodata \\
Tucana	                    &  $-$9.5  & (5.6 $\pm$ 1.6) $\times$ 10$^{5}$ &  880 $\pm$ 40 & 15.8 $\pm$ 3.6 &  130 $\pm$ 13 &  274 $\pm$ 40 & \nodata & $-$1.7 $\pm$ 0.2 & 0.30     & (4.3 $\pm$ 1.5) $\times$ 10$^{7}$ &   77 $\pm$ 34 & \nodata \\
And XVIII\tablenotemark{6}  &  $-$9.7  & (6.5 $\pm$ 3.1) $\times$ 10$^{5}$ & 1355 $\pm$ 88 &     \nodata    &   \nodata     &  363 $\pm$ 36 & \nodata & $-$1.8 $\pm$ 0.1 & \nodata     &       \nodata                    &    \nodata   & \nodata \\
\hline
\end{tabular}
\tablenotetext{}{See references in \S\,\ref{inventory} and Table~1 in \cite{wolf09}.}
\tablenotetext{1}{Errors in $r_c$ (core radius), $R_e$ (2D elliptical half light radius), and $r_{1/2}$ (3D deprojected half light radius) are set to 10\% (if not directly measured) for the dynamical mass estimates.}
\tablenotetext{2}{The total dynamical mass is calculated using the \cite{illingworth76} method, as described in \S\,\ref{meanvelocity}.  This method systematically underpredicts the total mass, as discussed in \S\S\,\ref{accuratemasses} and \ref{DMmasses}.}
\tablenotetext{3}{The mass at the half light radius is calculated using $M_{1/2}$ = 3$G^{-1}$$\langle$$\sigma_v^2$$\rangle$$r_{1/2}$ \citep{wolf09}, 
as described in \S\,\ref{accuratemasses}.}
\tablenotetext{4}{An error of 10\% in distance was assumed.}
\tablenotetext{5}{Given the uncertain measurement of the tip of the RGB, the distances to these satellites have been set to M31's mean distance, with a 10\% uncertainty.}
\tablenotetext{6}{An error of 0.5 mags was assumed for the integrated brightness of these galaxies.}
\tablenotetext{7}{For this galaxy, we assume $r_c$ $\sim$ $R_{e}$, with a large 25\% uncertainty.}
\label{table:params3}
\end{center}
\end{sidewaystable*}


\end{document}